\documentclass[12pt, a4paper]{article}

\usepackage{rotating}
\usepackage{amsmath}
\usepackage{mathtools}
\usepackage{amsthm}
\usepackage{amsfonts}
\usepackage[dvips]{epsfig}
\usepackage{graphicx}
\usepackage{psfrag}

\usepackage{amssymb}
\usepackage{cancel}
\usepackage{lscape}
\usepackage{cancel}
\usepackage{slashed}
\usepackage{lscape}
\usepackage{color}
\usepackage{cite}
\usepackage{url}
\usepackage{caption}
\usepackage{bbm}
\usepackage{bm}
\usepackage{subfig}
\usepackage{bbold}
\usepackage{caption}

\newcommand{\beq}{\begin{equation}}
\newcommand{\eeq}{\end{equation}}
\newcommand{\ga}{\lower.7ex\hbox{$\;\stackrel{\textstyle>}{\sim}\;$}}
\newcommand{\la}{\lower.7ex\hbox{$\;\stackrel{\textstyle<}{\sim}\;$}}

\makeatletter
\newcommand{\Cen}[2]{%
  \ifmeasuring@
    #2%
  \else
    \makebox[\ifcase\expandafter #1\maxcolumn@widths\fi]{$\displaystyle#2$}%
  \fi
}
\makeatother

\begin{document}

\begin{flushright}
{\tt KCL-PH-TH/2018-19}, {\tt CERN-TH/2018-123}  \\
{\tt UT-18-12, UMN-TH-3719/18, FTPI-MINN-18/10} \\

\end{flushright}

\vspace{0.5cm}
\begin{center}
{\bf {\large Uncertainties in WIMP Dark Matter Scattering Revisited}}
\end{center}

\vspace{0.05in}

\begin{center}{\large
{\bf John~Ellis}$^{a}$,
{\bf Natsumi Nagata}$^{b}$ and
{\bf Keith~A.~Olive}$^{c}$
}
\end{center}

\begin{center}
{\em $^a$Theoretical Particle Physics and Cosmology Group, Department of
  Physics, King's~College~London, London WC2R 2LS, United Kingdom;\\
Theoretical Physics Department, CERN, CH-1211 Geneva 23,
  Switzerland}\\
  {\em National Institute of Chemical Physics \& Biophysics, R{\" a}vala 10, 10143 Tallinn, Estonia; \\
Theoretical Physics Department, CERN, CH-1211 Geneva 23,
  Switzerland}\\[0.2cm]
  {\em $^b$Department of Physics, University of Tokyo, Bunkyo-ku, Tokyo
 113--0033, Japan}\\[0.2cm] 
{\em $^c$William I. Fine Theoretical Physics Institute, School of
 Physics and Astronomy, University of Minnesota, Minneapolis, MN 55455,
 USA}

\end{center}

\bigskip
\bigskip

\centerline{\bf ABSTRACT}

\noindent  
{\small We revisit the uncertainties in the calculation of spin-independent scattering
matrix elements for the scattering of WIMP dark matter particles on nuclear matter.
In addition to discussing the uncertainties due to limitations in our knowledge of the nucleonic
matrix elements of the light quark scalar densities $\langle N |{\bar u} u, {\bar d} d, {\bar s} s| N \rangle$,
we also discuss the importances of heavy quark scalar densities 
$\langle N |{\bar c} c, {\bar b} b, {\bar t} t| N \rangle$, and comment on uncertainties in quark mass ratios.
We analyze estimates of the light-quark densities made over the past decade using lattice calculations
and/or phenomenological inputs. We find an uncertainty in the combination
$\langle N |{\bar u} u + {\bar d} d | N \rangle$ that is larger than has been
assumed in some phenomenological analyses, and a range of $\langle N |{\bar s} s| N \rangle$
that is smaller but compatible with earlier estimates. We also analyze the importance of the ${\cal O}(\alpha_s^3)$
calculations of the heavy-quark matrix elements that are now available, which provide an important
refinement of the calculation of the spin-independent scattering cross section. We use for illustration a
benchmark CMSSM point in the focus-point region that is compatible with the limits from LHC and other searches.}

\vspace{0.2in}

\begin{flushleft}
May 2018
\end{flushleft}
\medskip
\noindent

\newpage

\section{Introduction}

Direct searches for the scattering of weakly-interacting massive particle (WIMP)
dark matter~\cite{GW} are proceeding apace, with regular increases in the experimental sensitivity \cite{lux16,XENON1T,pandax}
and plans for new experiments capable of probing spin-independent scattering
cross-sections $\sigma^p_{SI}$ approaching the neutrino floor \cite{Snowmass}. It is important that these important
experimental efforts be well served by theoretical efforts to minimize the uncertainties
in calculations of event rates within specific models of WIMP dark matter. These
include astrophysical uncertainties in the local density and velocity distribution
of the WIMPs, uncertainties in the accuracy with which effective interaction operator
coefficients can be calculated within a specific model, uncertainties in
the matrix elements of these operators in hadronic targets, and uncertainties in
nuclear structure effects. The focus of this paper is on the uncertainties in the
matrix elements for scattering on nucleon targets.

There are classes of two dimension-6 four-fermion interactions that yield
velocity-independent cross-sections for elastic WIMP-nucleon scattering, namely \cite{Falk:1998xj,efso}
\begin{equation}
{\cal L} \; = \sum_i  \; \alpha_{3i} \; {\bar \chi} \chi {\bar q}_i q_i +
\; \sum_i \; \alpha_{2i} \; {\bar \chi} \gamma_\mu \gamma_5 \chi {\bar q}_i \gamma^\mu \gamma_5 q_i \, ,
\label{dim6}
\end{equation}
where the sums are over the quark flavours $i$. Rates for the first set of interactions 
$\propto \alpha_{3i}$ are related to quark contributions to the nucleon mass: 
$m_{q_i} \langle N | {\bar q}_i q_i | N \rangle$ and are independent of the
nuclear spin, whereas rates for interactions
$\propto \alpha_{2i}$ are related to nucleonic matrix elements of axial currents in nucleons: 
$\langle N |{\bar q}_i \gamma_\mu \gamma_5 q_i | N \rangle$, which are related to quark
contributions to the nucleon spin. 
We discuss here the uncertainties
in the matrix elements of the spin-independent interactions $\propto \alpha_{3i}$,
which are relatively important, as we shall see. 
The uncertainties
in the spin-dependent interaction matrix elements $\propto \alpha_{2i}$ are relatively small,
as we discuss briefly towards the end of this paper.

Several approaches have been taken to estimating the $\langle N | {\bar q}_i q_i | N \rangle$
matrix elements. One of the first was to use octet baryon mass differences and SU(3) symmetry
to estimate the combination
$\sigma_0 \equiv \frac{1}{2} (m_u + m_d) \langle N | {\bar u} u + {\bar d} d - 2 {\bar s} s | N \rangle$,
together with data on low-energy $\pi N$ scattering to estimate the quantity
$\Sigma_{\pi N} \equiv \frac{1}{2} (m_u + m_d) \langle N | {\bar u} u + {\bar d} d | N \rangle$~\footnote{This 
quantity has also been estimated using data on pionic atoms~\cite{atoms}, with similar results, as we discuss later.}. As has 
been discussed in previous work, see, e.g.,~\cite{bot2,efso,eoss,eosv}, combining these estimates of $\sigma_0$ and $\Sigma_{\pi N}$ led to 
relatively large estimates for $\langle N | {\bar s} s | N \rangle$, though with large uncertainties. 

As we discuss below in some detail, in recent years a large effort has been put 
into lattice calculations, which have yielded a range of values of $\Sigma_{\pi N}$ and
relatively small estimates for $\langle N | {\bar s} s | N \rangle$. The corresponding values
of $\sigma_0$ may be similar to the estimates made using baryon masses and SU(3), but
some calculations correspond to significantly larger values of $\sigma_0$.
In parallel, there have been calculations using baryon chiral perturbation theory (B$\chi$PT)
that may lead to much larger values of $\sigma_0$, see, e.g.,~\cite{Alarcon:2012nr}, close to the data-based estimates of
$\Sigma_{\pi N}$, which may also correspond to relatively small values of $\langle N | {\bar s} s | N \rangle$.

In this paper we compile the lattice and other estimates of $\Sigma_{\pi N}$ and $\sigma_s \equiv m_s \langle N | {\bar s} s | N \rangle$
that have appeared over the past decade, and propose simple Gaussian representations of their values and
uncertainties. These may be useful for analyses of the constraints that direct searches for dark matter via
spin-independent scattering impose on specific models. Our combined estimate of $\Sigma_{\pi N}$ is similar
to values suggested previously, but with a larger uncertainty, while our combined estimate of $\sigma_s$ is
somewhat smaller than older estimates, though consistent with their uncertainties. 
We illustrate the results of our analysis with calculations of the spin-independent
dark matter scattering cross section $\sigma^p_{SI}$ at a specific benchmark point in the focus-point 
region \cite{fp} of the constrained minimal supersymmetric Standard Model (CMSSM) \cite{cmssm,azar}, 
noting that other supersymmetric parameter sets exhibit similar trends. As we discuss, the uncertainties in $\sigma^p_{SI}$ related
to light quark masses and $\langle N | {\bar u} u | N \rangle /  \langle N | {\bar d} d | N \rangle$ are significantly
smaller than those associated with $\sigma_s$. We also discuss the uncertainties in $\sigma^p_{SI}$ associated with the
heavy quark matrix elements $\langle N | {\bar c} c, {\bar b} b, {\bar t} t | N \rangle$. The $b$ and $t$ quarks are
sufficiently heavy that a perturbative treatment of their hadronic matrix elements is appropriate, but this is not
so evident for the $c$ quark. Some lattice and other numerical estimates of $\sigma_c \equiv m_c \langle N | {\bar c} c | N \rangle$ are
available, and span a wide range that straddles the perturbative estimate. If the ${\cal O}(\alpha_s^3)$ perturbative estimates are
used for all the heavy-quark matrix elements, as we advocate,
the corresponding uncertainties in $\sigma^p_{SI}$ are small, but if the full range of
numerical estimates of $\sigma_c$ is considered the associated uncertainty is comparable to that associated with $\sigma_s$.

The layout of this paper is as follows. In Section~\ref{sec:SI} we review the
strong-interaction quantities entering the calculation of the spin-independent 
cross-section $\sigma^p_{SI}$. Inputs to the calculation of $\sigma^p_{SI}$ are discussed in Subsection~\ref{sec:inputs}, 
the individual uncertainties in the matrix elements of the densities of the light quarks $u, d, s$
are discussed in Subsection~\ref{sec:piN}, and their propagation into the calculation of
$\sigma^p_{SI}$ are discussed in Subsection~\ref{sec:densityMEs}. Subsection~\ref{sec:HQ} is dedicated to a discussion
of the matrix elements of the heavy quarks $c, b, t$.  Finally, Section~\ref{sec:SD}
contains a brief discussion of the uncertainties in the calculation of the spin-dependent 
cross-section $\sigma_{SD}$, and our conclusions are summarized in Section~\ref{sec:conx}.

\section{Spin-Independent WIMP-Nucleon Scattering}
\label{sec:SI}

\subsection{Inputs to the Matrix Element Calculation}
\label{sec:inputs}

At zero momentum transfer, and neglecting nuclear structure effects, the spin-independent
cross-section for the elastic scattering of a generic WIMP on a nucleus with charge $Z$
and atomic number $A$ can be written as
\cite{GW,kg,bfg,Ellis:1992ka,dn,GJK,Falk:1998xj, Hisano:2010ct}
\begin{equation}
\sigma^{Z,A}_{SI} \; = \; \frac{4 m_r^2}{\pi} \left[ Z f_p + (A - Z) f_n \right]^2 \, ,
\label{sigmaSI}
\end{equation}
where $m_r$ is the reduced WIMP mass,
\begin{equation}
\frac{f_N}{m_N} \; = \; \sum_q f^N_{T_q} \frac{\alpha_{3q}}{m_q}
\label{fN}
\end{equation}
for $N = p$ or $n$, and the quantities $f^N_{T_q}$ are defined by
\begin{equation}
m_N f^N_{T_q} \; \equiv \; \langle N | m_q {\bar q} q | N \rangle \; \equiv \; \sigma_q \; \equiv \; m_q B^N_q \, .
\label{fTq}
\end{equation}
We recall that the quantities $f^N_{T_q} \; (\sigma_q)$ are independent of renormalization scheme and scale,
whereas the quantities $m_q$ and $B^N_q$ appearing have cancelling scheme and scale
dependences.

The contributions of the heavy quarks $q = c, b, t$ have often been treated by integrating them out
and replacing them by the one-loop contributions due to scattering off gluons \cite{SVZ}, so that
\begin{equation}
\frac{f_N}{m_N} \; = \; \sum_{q = u, d, s} f^N_{T_q} \frac{\alpha_{3q}}{m_q}
+ \frac{2}{27} f^N_{T_G} \sum_{q = c, b, t} \frac{\alpha_{3q}}{m_q} \, ,
\label{f3}
\end{equation}
where 
\begin{equation}
f^N_{T_G} \; = \; 1 - \sum_{q = u,d,s} f^N_{T_q} \, .
\label{fNTG}
\end{equation}
In our subsequent analysis, this is the first approach we use to calculate $\sigma^{p, n}_{SI}$.

However, as discussed later in more detail, there are by now a number of lattice calculations of 
$f^N_{T_c} = m_c \langle N | {\bar c} c | N \rangle / m_N = \sigma_c/m_N$, so for comparison we also estimate $\sigma^p_{SI}$
using the one-loop 4-flavour versions of (\ref{f3}) and (\ref{fNTG}), where
\begin{equation}
\sum_{q = u,d,s} \to \sum_{q = u,d,s,c} \, , \qquad \sum_{q = c,b,t} \to \sum_{q = b,t} \, , \qquad \frac{2}{27} \to \frac{2}{25} \, ,
\label{f4}
\end{equation}
together with a numerical estimate of $f^N_{T_c}$ that is based (mainly) on lattice calculations.

As we also discuss later, there are also calculations of $f^N_{T_c}, f^N_{T_b}$ and $f^N_{T_t}$ to ${\cal O}(\alpha_s^3)$ in perturbative QCD.
These perturbative calculations are expected to be very reliable for $f^N_{T_b}$ and $f^N_{T_t}$, perhaps less so for $f^N_{T_c}$.
Therefore, we also estimate $\sigma^p_{SI}$ using these calculations in the full six-flavour formula (\ref{fN}), for comparison with
the three-quark formula (\ref{f3}) and the four-quark formula (\ref{f4}) evaluated using the available numerical estimates of $f^N_{T_c}$.
As we discuss later, we consider the ${\cal O}(\alpha_s^3)$ six-flavour approach to be the best available at the present time.

In order to evaluate (\ref{f3}) we need estimates of the matrix elements
$\langle N| {\bar u}u, {\bar d} d, {\bar s} s| N \rangle$, for which isospin invariance 
ensures that $\langle p| {\bar u}u| p \rangle = \langle n| {\bar d}d| n \rangle = B_u^p$,
$\langle p| {\bar d} d| p \rangle = \langle n| {\bar u} u| n \rangle = B_d^p$ and $\langle p| {\bar s} s| p \rangle =
\langle n| {\bar s} s| n \rangle = B_s^p$. An expression for one combination of these quantities is provided by the
pion-nucleon $\sigma$ term
\begin{equation}
\Sigma_{\pi N} \; = \; \frac{1}{2} (m_u + m_d) \left( B_u^p + B_d^p \right) \, ,
\label{piNSigma}
\end{equation}
which may be extracted phenomenologically from data on low-energy $\pi$-nucleon scattering
or on pionic atoms~\cite{atoms}. In order to determine $B_s^p$, (\ref{piNSigma}) has often been used in combination with the quantity
\begin{equation}
\sigma_0 \; = \; \frac{1}{2} (m_u + m_d) \left( B_u^p + B_d^p - 2 B_s^p\right) \, ,
\label{sigma0}
\end{equation}
which can be extracted phenomenologically from the octet baryon mass splittings,
taking into account corrections that can be calculated in baryonic chiral perturbation theory (B$\chi$PT).
One then has
\beq
\sigma_s  = m_s B^p_s = \frac{m_s}{m_u+m_d} (\Sigma_{\pi N} -\sigma_0)\, ,
\label{sigmas}
\eeq
which is often parameterized by
\beq
y = 1 - \frac{\sigma_{0}}{\Sigma^p_{\pi N}} = \frac{2 B^p_s}{B^p_u+B^p_d} \, .
\eeq
However, alternatives to these phenomenological estimates are now provided by the many lattice
calculations that are now available, as we discuss in Section~\ref{sec:piN} below. 

In order to evaluate $\Sigma_{\pi N}$, we also need values for the ratios of 
quark masses $m_u/m_d$ and $m_s/m_d$. In the past~\cite{eosv,mc12}, we have used the estimates
\begin{equation}
\frac{m_u}{m_d} \; = \; 0.553(43), \quad \frac{m_s}{m_d} \; = \; 18.9(8)
\label{ratios}
\end{equation}
from~\cite{Leutwyler:1996qg}, whereas the Particle Data Group (PDG)
now quotes the following lattice estimates \cite{Patrignani:2016xqp}:
\begin{equation}
\frac{m_u}{m_d} \; = \; 0.46 (5), \quad \frac{2 m_s}{m_u + m_d} \; = \; 27.5 (3), \quad \to \quad \frac{m_s}{m_d}  \; = \; 20.1 (8) \, .
\label{PDGratios}
\end{equation}
The PDG lattice review also quotes the following absolute values of the
light quark masses in the $\overline{\rm MS}$
scheme at a renormalization scale of 2~GeV:
\begin{equation}
m_u \; = \; 2.15(15)~{\rm MeV}, \quad m_d \; = \; 4.70(20)~{\rm MeV}, \quad m_s \; = \; 93.5 \pm 2~{\rm MeV} \, .
\label{udsmasses}
\end{equation}
In contrast, in its Summary Tables the PDG quotes the broader ranges \cite{Patrignani:2016xqp}
\begin{equation}
\frac{m_u}{m_d} \; = \; 0.38 \; {\rm to} \; 0.58, \quad \frac{2 m_s}{m_u + m_d} \; = \; 27.3 (7), \quad \frac{m_s}{m_d}  \; = \; 17 \; {\rm to} \; 22
\label{PDGSummaryratios}
\end{equation}
and
\begin{equation}
m_u \; = \; 2.2^{+0.6}_{-0.4}~{\rm MeV}, \quad m_d \; = \; 4.7^{+0.5}_{-0.4}~{\rm MeV}, \quad m_s \; = \; 96^{+8}_{-4}~{\rm MeV} \, .
\label{udsSummarymasses}
\end{equation}
As we discuss below, the resulting elastic scattering cross sections we study are relatively insensitive to the
values of the quark mass ratios and are completely insensitive to the absolute value of the quark mass (e.g. $m_d$). 
For definiteness, we use (\ref{udsmasses}) for the value of the strange quark mass and (\ref{PDGratios}) for the values of the quark mass ratios.

The quantities $B_u + B_d$ and $B_s$ discussed above suffice to calculate the matrix elements
for scattering off nuclei with equal numbers of protons and neutrons, but additional information is
required to calculate the difference between the cross sections for scattering off protons and neutrons, 
or for the scattering off nuclei with general values of $(A, Z)$, as seen in (\ref{sigmaSI}).

Another combination of the $B_q^p$ was calculated~\cite{Cheng:1988im} using octet baryon
masses in the relation:
\begin{equation}
z \; \equiv \; \frac{B_u^p - B_s^p}{B_d^p - B_s^p} \; = \; \frac{m_{\Xi^0} + m_{\Xi^-} - m_p - m_n}{m_{\Sigma^+} + m_{\Sigma^-} - m_p - m_n} \; = \; 1.49 \, .
\label{zratio}
\end{equation}
The uncertainty associated with measurements of the octet baryon masses is very small,
but the accuracy of the octet mass formula (\ref{zratio}) is subject to question~\cite{Crivellin:2013ipa}.

Alternatively, one can calculate $z$ from the ratio
\begin{equation} \label{eqn:BdBu}
z \; = \; \frac{2 - (1 + \frac{B_d^p}{B_u^p} ) y }{(2 - y) \frac{B_d^p}{B_u^p} - y }
\end{equation}
once $B_u^p/B_d^p$ and the value of $y$ are known. For example, the ratio $B_u^p/B_d^p$ can be calculated from the QCD
contribution to the proton-neutron mass difference:
\begin{equation}
m_p - m_n|_{QCD} \; = \; (m_u - m_d) \left( B_u^p - B_d^p \right) \, ,
\label{pnmassdifference}
\end{equation}
in combination with (\ref{piNSigma}):
\begin{equation}
\frac{\left(\frac{m_u}{m_d} - 1\right)}{\left(\frac{m_u}{m_d} + 1 \right)} \, \frac{\left( 1 - \frac{B_d^p}{B_u^p} \right)}{ \left( 1 + \frac{B_d^p}{B_u^p} \right)} \; = \;  \frac{(m_p - m_n )|_{QCD}}{2 \Sigma_{\pi N}} \, ,
\label{Bu-Bd}
\end{equation}
using the value (\ref{PDGratios}) of $m_u/m_d$.
The measured mass difference $m_p - m_n = - 1.29$~MeV (with negligible uncertainty)
and the electromagnetic contribution is estimated to be $m_p - m_n|_{QED} = 1.04 (11)$~MeV \cite{Thomas:2014dxa},
leading to $m_p - m_n|_{QCD} = - 2.33 (11)$~MeV. Inserting the central value of $m_u/m_d$ from (\ref{PDGratios})
and the central value of $\Sigma_{\pi N}$ from (\ref{95percent}) below into (\ref{Bu-Bd}) to evaluate
$B^p_d/B^p_u$, and then using (\ref{eqn:BdBu}) with $y$ given by the central values of $\Sigma_{\pi N}$ and $\sigma_s$
in (\ref{95percent}) and (\ref{95percentss}) below, respectively, we estimate
\begin{equation}
z \; = \; 1.16 \, .
\label{secondz}
\end{equation}
A similar value $z = 1.258 (81)$ was found independently in a lattice calculation in~\cite{Bali:2016lvx}.
Conservatively, we consider the range $1 < z < 2$ in our subsequent estimates.

\subsection{Uncertainties in $\Sigma_{\pi N}$, $\sigma_0$ and $\sigma_s$}
\label{sec:piN}

\begin{figure}[ht!]
\centering
   \subfloat{\scalebox{0.35}{\includegraphics{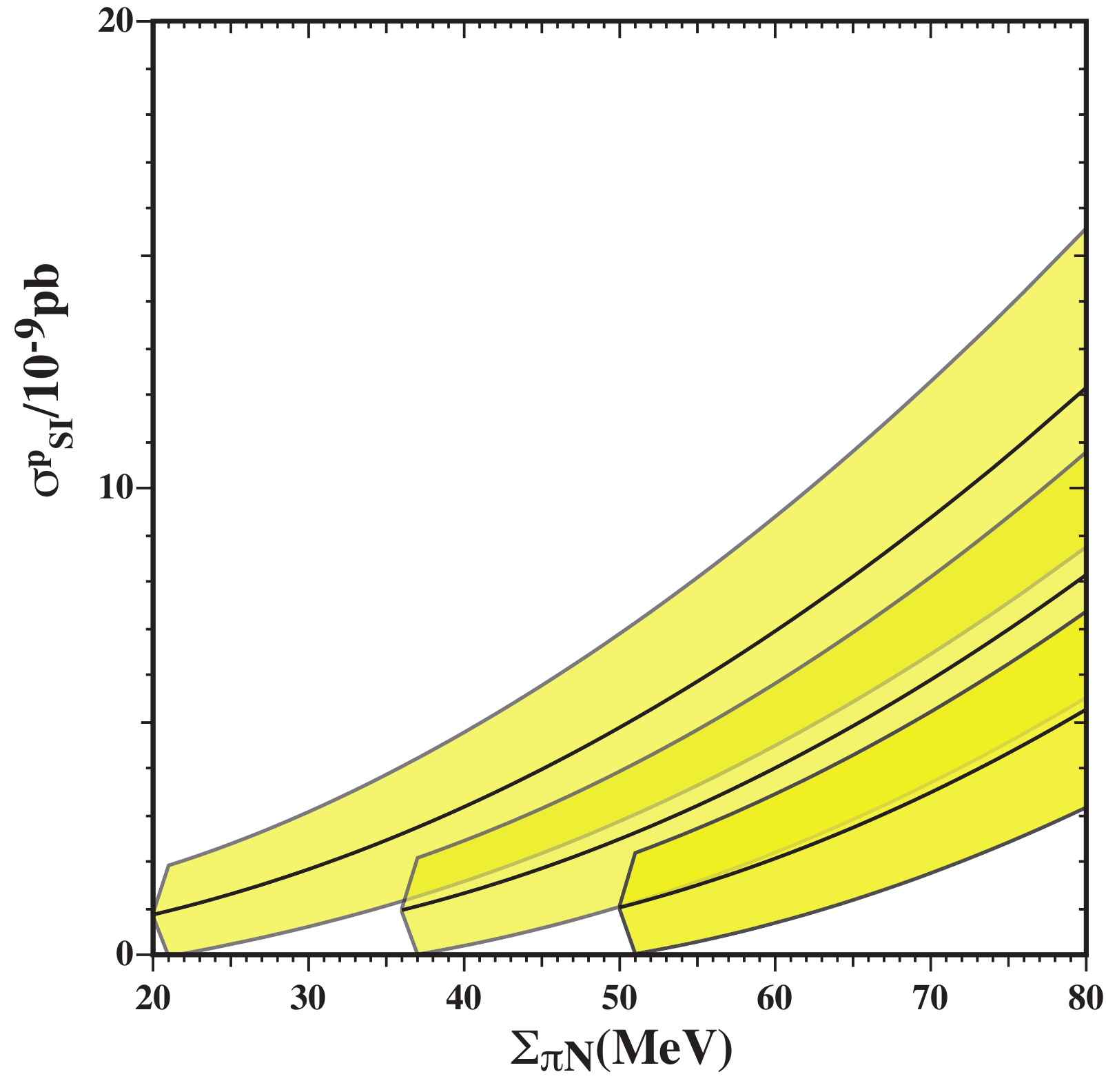}}}
\hskip .2in
    \subfloat{\scalebox{0.35}{\includegraphics{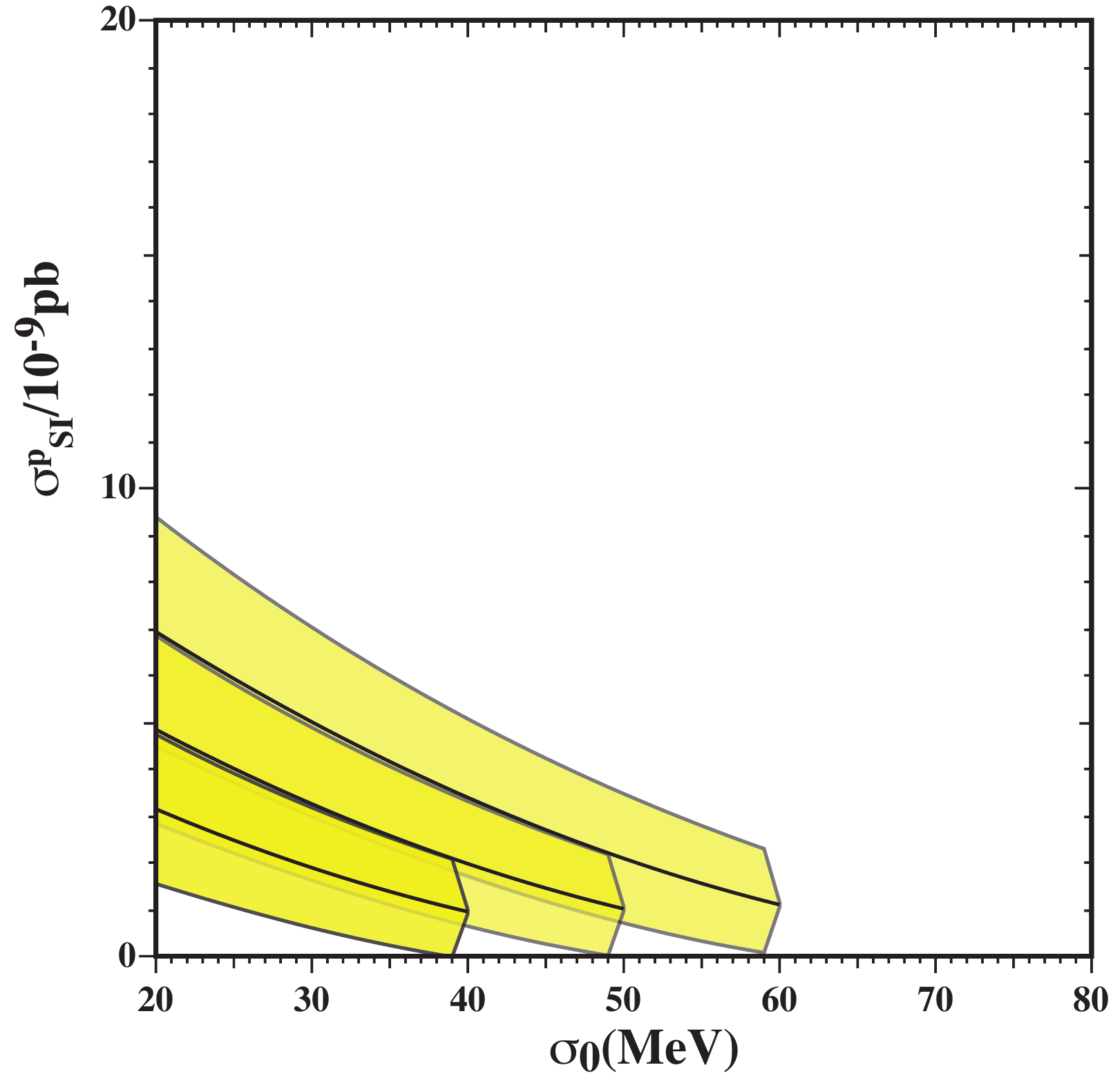}}}
    \caption{\em Left: $\sigma^p_{SI}$ vs $\Sigma_{\pi N}$ for $\sigma_0 = 20, 36, 50$ MeV. 
    Right: $\sigma^p_{SI}$ vs $\sigma_0$ for  $\Sigma_{\pi N} =40, 50, 60$ MeV.
    The color bands show the 1-$\sigma$ uncertainty in the elastic cross section
    calculated using the three-flavour expression (\protect\ref{f3}). 
        }
    \label{fig:sigma0sigmas}
\end{figure}

The most important uncertainties in calculations of the spin-independent WIMP-nuclear
scattering matrix elements are those in $\Sigma_{\pi N}$ and $\sigma_s$.
In the past, $\sigma_s$ has often been evaluated by combining phenomenological estimates of 
$\Sigma_{\pi N}$ (\ref{piNSigma}) and $\sigma_0$ (\ref{sigma0}) using (\ref{sigmas}).
The uncertainties in $\Sigma_{\pi N}$ and $\sigma_0$ translate into significant uncertainties in the spin-independent cross section
\cite{efso,bot2,eoss,eosv}~\footnote{We describe the propagation of these uncertainties in Section \ref{sec:densityMEs}.},
as illustrated in Fig.~\ref{fig:sigma0sigmas}, where the three-flavour expression (\ref{f3}) has been used. 
Here and in the analysis that follows, we use a representative point in the CMSSM consistent with the limits from LHC and other searches \cite{azar}, namely
with $m_{1/2} = 3000$ GeV, $m_0 = 8200$ GeV, 
$A_0 = 0$ GeV, $\tan \beta = 10$, and $\mu > 0$. At this point, the LSP is mainly a
Higgsino with mass $\simeq 1.1$ TeV whose relic density matches that determined by CMB
experiments \cite{cmb}. We have verified that similar trends in the dependences on $\Sigma_{\pi N}$ and $\sigma_0$
arise at other representative points, namely a stop-coannihilation point and s-channel A/H funnel point, though the values of the elastic cross section is very different
for these points.

The left panel of Fig.~\ref{fig:sigma0sigmas} shows the values of $\sigma^p_{SI}$ obtained as a
function of $\Sigma_{\pi N}$ for three indicative values of $\sigma_0 = 20, 36, 50$ MeV, and the
right panel of Fig.~\ref{fig:sigma0sigmas} shows the values of $\sigma^p_{SI}$ obtained as a
function of $\Sigma_{0}$ for the three indicative values $\Sigma_{\pi N} =40, 50, 60$ MeV.
In the two cases, representative uncertainties of 7 MeV were assumed in $\sigma_0$ and
$\Sigma_{\pi N}$, respectively. In making these plots, we have assumed that $B_s \ge 0$
and hence imposed the restriction $\Sigma_{\pi N} \ge \sigma_0$.  For the indicative values 
$\Sigma_{\pi N} = 50 \pm 7$ MeV and $\sigma_0 = 36 \pm 7 $ MeV, we find 
$\sigma^p_{SI} = (2.5 \pm 1.5) \times 10^{-9}$ pb using the three-flavour formula (\ref{f3}).

`Legacy' values of $\Sigma_{\pi N}$ and $\sigma_0$ quoted in~\cite{eosv} were $\Sigma_{\pi N} = 64 (8)$~MeV from
$\pi$-N scattering and $\sigma_0 = 36 (7)$~MeV from octet baryon mass differences~\cite{Borasoy,Gasser:1990ce,Knecht:1999dp,Sainio:2001bq,pavan}.
Subsequently, since lattice calculations have tended to yield lower values of $\Sigma_{\pi N} \gtrsim 40$~MeV,
and the MasterCode collaboration has been using the `compromise' value 
$\Sigma_{\pi N} = 50 (7)$~MeV when using the experimental limits on spin-independent dark matter scattering on nuclei
in global fits to supersymmetric models (see, e.g., \cite{mc12}). In combination
with $\sigma_0 = 36 (7)$~MeV~\cite{Borasoy,Gasser:1990ce,Knecht:1999dp,Sainio:2001bq,pavan}, 
this yields $\sigma_s = 192(136)$~MeV. Another global fitting group, the GAMBIT
Collaboration (see, e.g., \cite{gambit}), has, on the other hand, been using the smaller value $\sigma_s = 43 (8)$~MeV,
which is based on a compilation of lattice data made in 2011~\cite{Lin:2011ab}, together with a larger value of 
$\Sigma_{\pi N} = 58 (9)$~MeV. This combination corresponds to $\sigma_0 = 55$ MeV, considerably
larger than the estimate from octet baryon mass differences~\cite{Borasoy}, but within the range argued in~\cite{Alarcon:2012nr}
to be consistent with B$\chi$PT.

Here we revisit the uncertainties in $\Sigma_{\pi N}$ and $\sigma_s$ based on the 
considerable effort during the last decade made since~\cite{eosv},
using lattice and other techniques, to determine 
$\Sigma_{\pi N}$ and $\sigma_s$ \cite{joel}. Although most of these recent values have been obtained from 
lattice calculations, many have been based on the phenomenology of low-energy $\pi$-nucleon interactions, 
and some have made extensive use of chiral perturbation theory, often in combination with lattice techniques.
As already commented in~\cite{1602.07688}, and discussed in more detail below, there is tension between
these various estimates, and the uncertainties are not purely statistical.

\begin{figure}[ht!]
\centering
   \subfloat{\scalebox{0.5}{\includegraphics{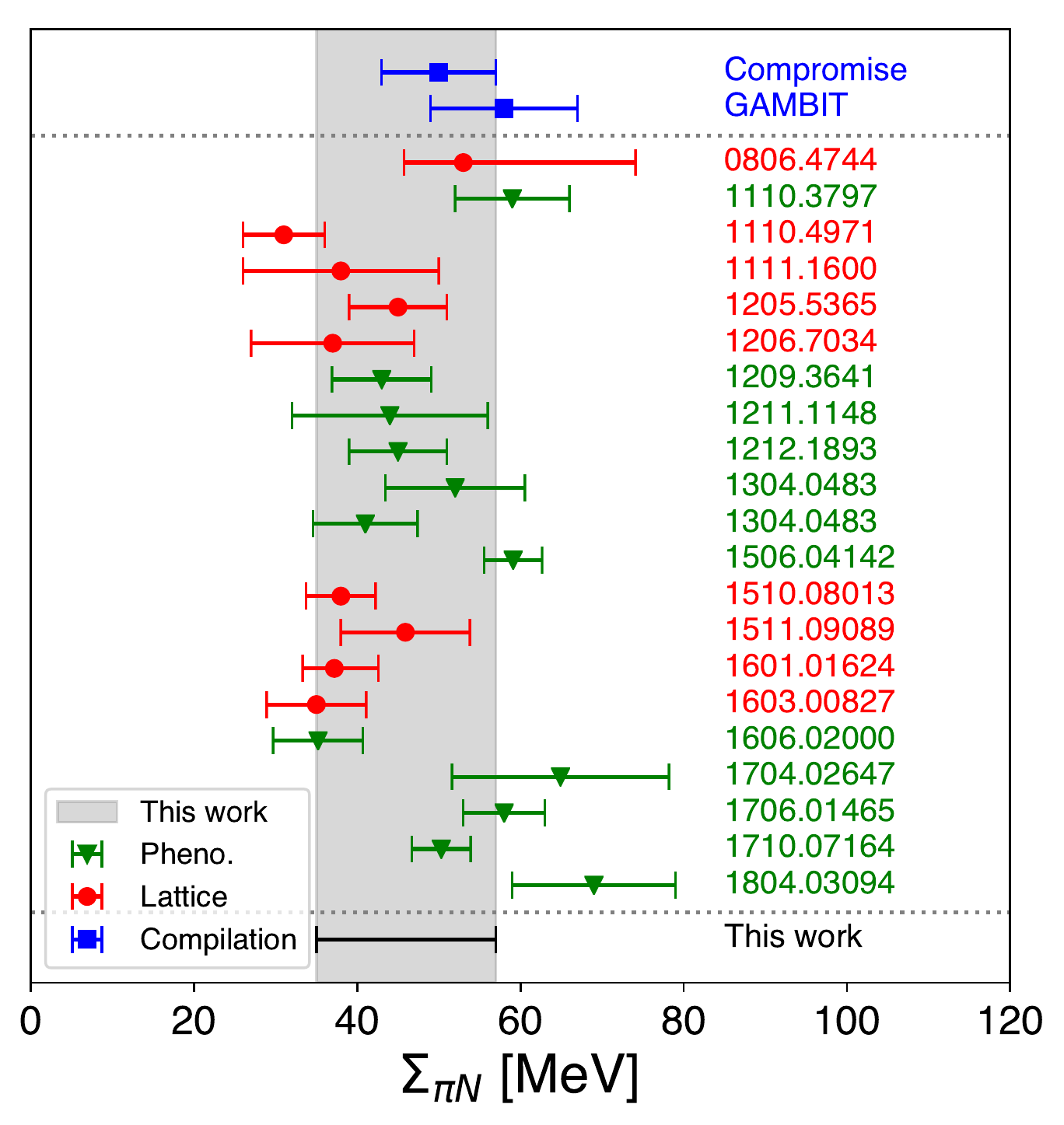}}}
    \hspace{1cm}
    \subfloat{\scalebox{0.5}{\includegraphics{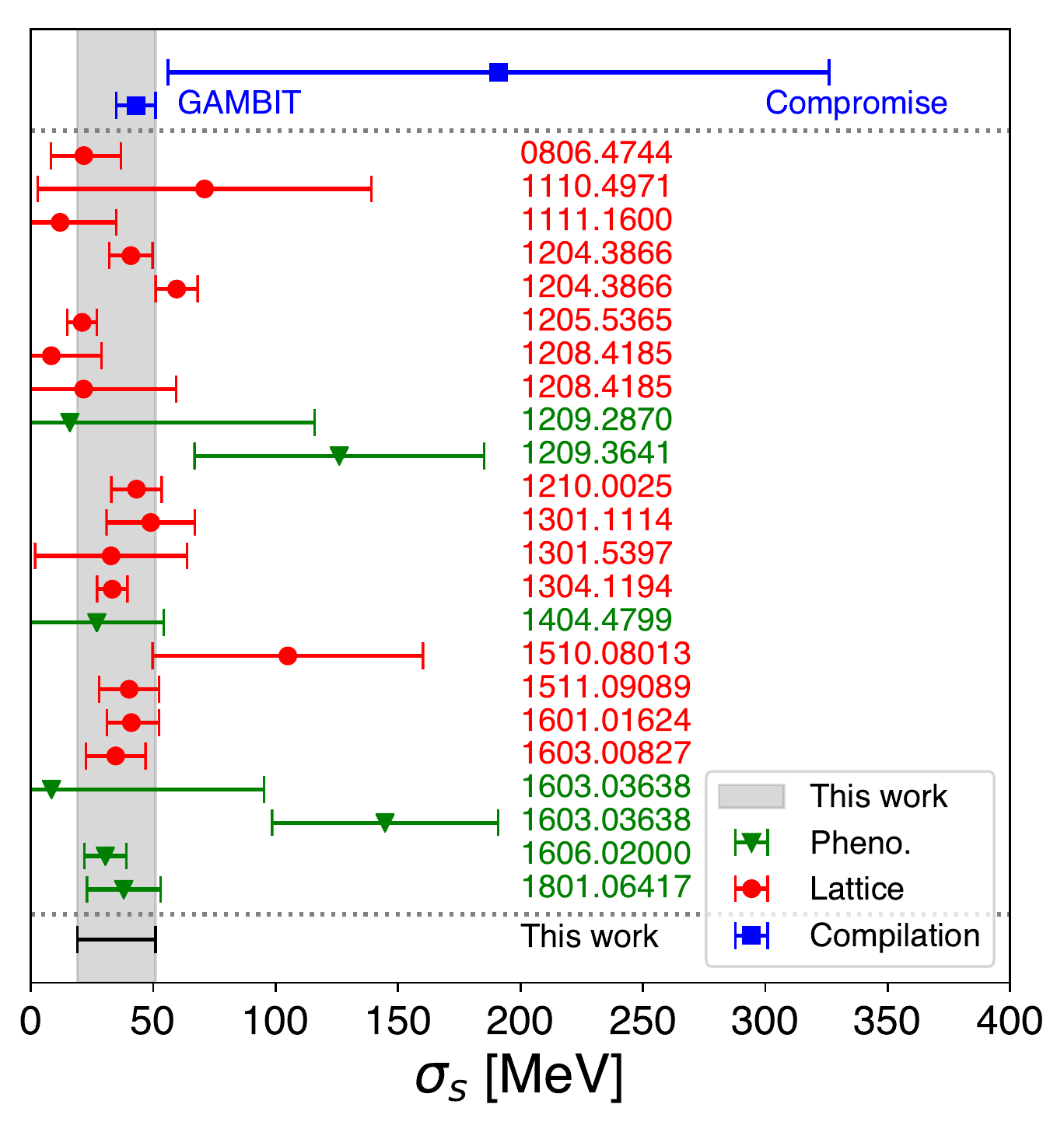}}}
    \caption{\em Left panel: Recent values of $\Sigma_{\pi N}$. Right panel: Recent values of $\sigma_s$. 
    Calculations based mainly on lattice calculations are indicated in red, and those relying more on
    phenomenological inputs are indicated in green.
    We also show the estimates made in previous compilations~\cite{mc12,gambit} (blue), 
    and the values we estimate now on the basis of our new compilation (bottom line and vertical grey bands).
        }
    \label{fig:spiNfigssfig}
\end{figure}

The left panel of Fig.~\ref{fig:spiNfigssfig} displays all the estimates of $\Sigma_{\pi N}$
that we use, and the right panel displays all the estimates of $\sigma_s$ that are included in our analysis.
We have tried to make a complete selection of all the determinations of these quantities that have not
been superseded by later analyses by strongly-overlapping research groups. In each case, we have indicated
by colour coding the primary phenomenological technique used in the calculation, and we have also indicated the
corresponding arXiv reference number. We also indicate by shaded bands in Fig.~\ref{fig:spiNfigssfig} the estimates of
$\Sigma_{\pi N}$ and $\sigma_s$ that we make on the basis of this new compilation, using the prescription that we describe
below. More details of the determinations we use, including their numerical values, are
given in Table~\ref{tab:values}~\footnote{We apologize in advance to authors whose work we have overlooked or
misrepresented in compiling this Table, and welcome suggestions for its completion and improvement.}.

\begin{table}[ht!]
\caption{Estimates of $\Sigma_{\pi N}$ and $\sigma_s$. }
\label{tab:values}
\vspace{-0.5cm}
\begin{center}
\scriptsize
\begin{tabular}{|c||c|c||c|c||c|}
\hline
Reference & $\Sigma_{\pi N}$ & Uncertainties & $\sigma_s$ & Uncertainties & Method \\
\hline
\hline
\cite{mc12} & 50 & 7 & 191 & 135 & Compilation \\
\cite{gambit} & 58 & 9 & 43 & 8 & Compilation \\
\hline
\hline
\cite{Ohki:2008ff}  & 53 &  $ 2^{+21}_{-7}$ &  21.7 & ${}^{+15.1}_{-13.4}$ &  Lattice \\
\hline
\cite{Alarcon:2011zs} & 59 & 7 & & & B$\chi$PT, $\pi$ atoms \\
\hline
\cite{Horsley:2011wr} & 31 & $3 \pm 4$ & 71 & $34 \pm 59$ & Lattice \\
\hline
\cite{Bali:2011ks} & 38 & 12 & 12 & $^{+23}_{-16}$ & Lattice  \\
\hline
\cite{Freeman:2012ry}  & &  & 40.9 & $7.5 \pm 4.7$ & Lattice  \\
& & & 59.6 & $5.1 \pm 6.9$ & Lattice  \\
\hline
\cite{Shanahan:2012wh} & 45 & 6 & 21 & 6 & Lattice  \\
 \hline
\cite{Bali:2012qs} & 37 & $8 \pm 6$ & & & Lattice \\
\hline
\cite{Oksuzian:2012rzb}  &  & & 8.4 & $14.1 \pm 15.0$ & Lattice  \\
 & & & 21.6 & $27.2 \pm 26.3$ & Lattice  \\
\hline
\cite{Alarcon:2012nr} & & & 16 & $80 \pm 60$ & B$\chi$PT \\
\hline
\cite{Ren:2012aj}  & 43 & $1\pm 6$ & 126 & $24 \pm 54$ & Lattice/B$\chi$PT  \\
\hline 
\cite{Engelhardt:2012gd} & & & 43.2 & 10.3 & Lattice  \\
\hline
\cite{Stahov:2012ca}  & 44 & 12 & & & $\pi$N  scattering\\
\hline
  \cite{Chen:2012nx} & 45 & 6 & & & $\pi$N scattering \\
\hline
\cite{Junnarkar:2013ac} & & & 49 & $10 \pm 15$ & Lattice  \\
\hline
\cite{Jung:2012rz} & & & 32.8& 31.0 & Lattice  \\
\hline
\cite{Alvarez-Ruso:2013fza}  & 52 & $3 \pm 8$ & & & Lattice/B$\chi$PT  \\
& 41 & $ 5 \pm 4$ &&& Lattice/B$\chi$PT \\
\hline
 \cite{Gong:2013vja} & & & 33.3 & 6.2 & Lattice/B$\chi$PT \\
\hline
\cite{Ren:2014vea} &  &  & 27 & $27 \pm 4$ & Lattice/B$\chi$PT \\
\hline
\cite{Hoferichter:2015dsa}  & 59.1 & $1.9 \pm 3$ &&& $\pi$ atoms \\
\hline
\cite{Durr:2015dna} & 38 & $ 3 \pm 3$ & 105 & $41 \pm 37$ & Lattice \\
\hline
\cite{Yang:2015uis} & 45.9 & $ 7.4 \pm 2.8$ & 40.2 & $11.7 \pm 3.5$ & Lattice \\
\hline
\cite{Abdel-Rehim:2016won}  & 37.2 & $ 2.6^{+4.7}_{-2.9}$ & 41.1 & $8.2^{+7.8}_{-5.8}$ & Lattice \\
\hline
\cite{Bali:2016lvx}  & 35 & 6.1 & 34.7 & 12.2 & Lattice \\
\hline
\cite{Yao:2016vbz}  & & & 8.5 & $4.4 \pm 86.6$ & $\pi$ atoms, $\pi$N scattering \\
& & & 144.7 & $4.6 \pm 45.9$ & $\pi$ atoms, $\pi$N scattering \\
\hline
\cite{Duan:2016rkr}  & 35.2 & $ 5.5 $ &30.5 & 8.5  & B$\chi$PT \\
\hline
\cite{Alexandrou:2017xwd}  & 64.9 & $ 1.5 \pm 13.2$ & & & Lattice/B$\chi$PT \\
\hline
\cite{RuizdeElvira:2017stg}  & 58 & 5 & & & $\pi$N scattering \\
\hline
\cite{Ling:2017jyz} & 50.3 & $1.2 \pm 3.4$ & & & Lattice/B$\chi$PT \\
\hline
\cite{Lutz:2018cqo} & 48 & & 38 & 15 & Lattice/B$\chi$PT \\
\hline
\cite{Fernando:2018jrz} & 69 & 10 & & &B$\chi$PT  \\
\hline
\hline
{\bf This work} & {\bf 46} & {\bf 11} & {\bf 35} & {\bf 16} & {\bf New compilation} \\
\hline
\end{tabular}
\end{center}
\caption*{\it Estimates of $\Sigma_{\pi N}$ and $\sigma_s$ (in MeV units). The first two lines are from
previous compilations. The following lines are from recent determinations and, where two
errors are quoted, the first is statistical and the second systematic. As indicated, most of the determinations
are based on lattice calculations, many use baryon chiral perturbation theory (B$\chi$PT), three use
data on pionic atoms ($\pi$ atoms) and five use low-energy $\pi$N scattering data.
The last line is our new compilation.}
\end{table}

\begin{figure}[t!]
\centering
   \subfloat{\scalebox{.7}{\includegraphics{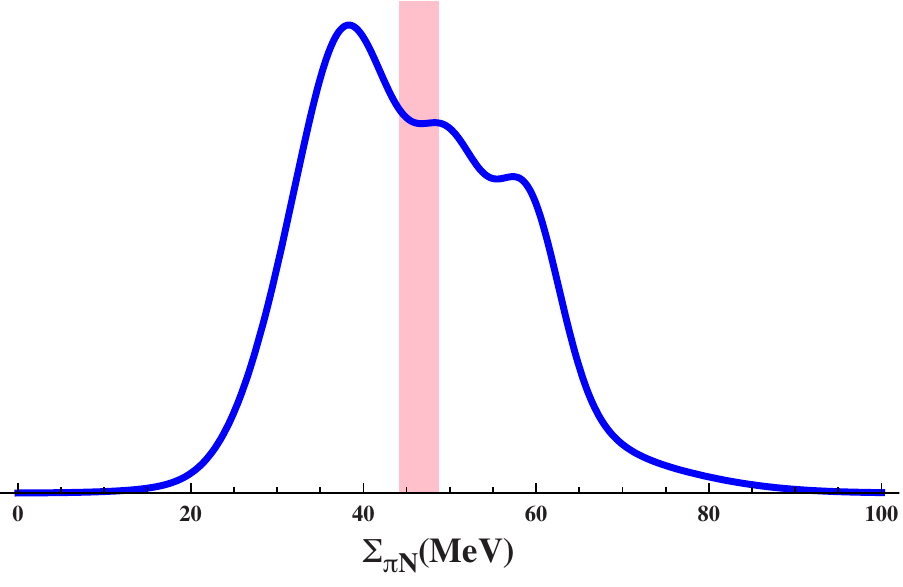}}}
    \hspace{1cm}
    \subfloat{\scalebox{.7}{\includegraphics{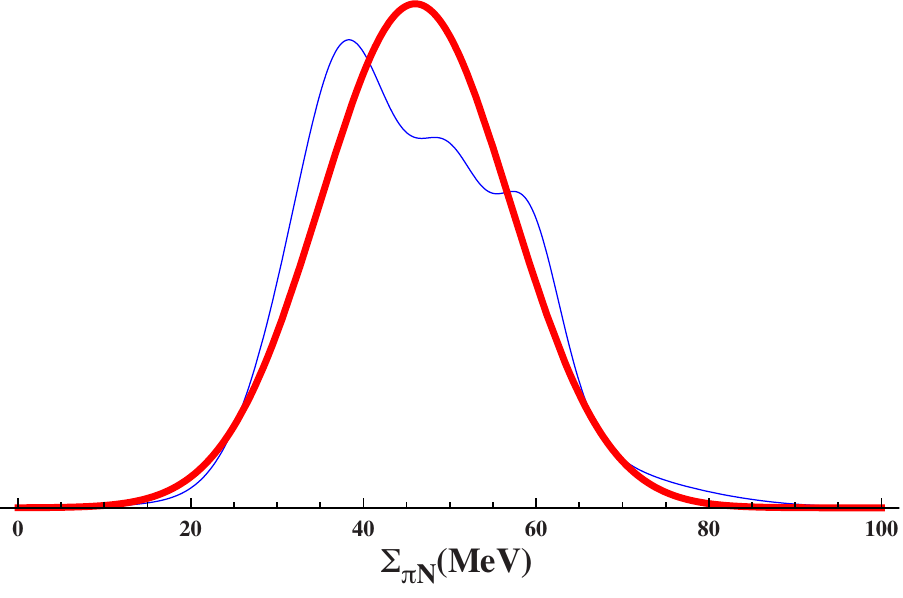}}}
    \caption{\em Left panel: Ideogram combining all values of $\Sigma_{\pi N}$ from Table \ref{tab:values}.
    The vertical (pink) bar corresponds to the estimate (\protect\ref{rescaled}).
    Right panel: The same ideogram compared with a (red) Gaussian centred at $\Sigma_{\pi N} = 46$~MeV with error $\sigma = 11$~MeV.}
    \label{fig:ideograms}
\end{figure}

We now discuss the combinations of these estimates using the procedures adopted by the Particle Data Group (PDG) 
in cases where the uncertainties are not simply statistical~\cite{Patrignani:2016xqp}.
Assuming uncorrelated Gaussian probability distributions for each of the estimates of $\Sigma_{\pi N}$ shown in the left panel of
Fig.~\ref{fig:spiNfigssfig}, we first construct the ideogram\footnote{The ideogram is constructed using the prescription of the 
PDG~\cite{Patrignani:2016xqp}, and is a sum of Gaussians for each measurement with an area normalized to be $1/\sigma_i$ where
$\sigma_i$ is the uncertainty in the measurement.}
 shown in the left panel of Fig.~\ref{fig:ideograms}.
As can be discerned from Fig.~\ref{fig:spiNfigssfig}, the values of $\Sigma_{\pi N}$ are broadly distributed 
between 40-60 MeV, and the ideogram exhibits 3 minor peaks, slightly favoring the lower part of the 
range. 

A naive weighted mean of all 21 determinations of $\Sigma_{\pi N}$ yields
\begin{equation}
{\rm Naive:} \qquad \Sigma_{\pi N} \; = \; 46.1 \pm 1.3 \; {\rm MeV} \, ,
\label{naive}
\end{equation}
where we have combined statistical and systematic uncertainties in quadrature and centred asymmetric errors. 
It is clear, however,  that this naive estimate would be a poor representation of the ideogram.
One option proposed by the PDG under such circumstances is to rescale the error so that the $\chi^2$/d.o.f.$ = 1$. In this case, the required
renormalization factor is 1.7, yielding 
\begin{equation}
{\rm Rescaled:} \qquad \Sigma_{\pi N} \; = \; 46.1 \pm 2.2 \; {\rm MeV} \, .
\label{rescaled}
\end{equation}
However, this would also be a poor representation of the ideogram, in view of its serrated ridge top
that is broader than the rescaled distribution (\ref{rescaled})~\footnote{This feature may reflect the existence
of unidentified systematic uncertainties that affect different lattice methods and B$\chi$PT approaches in
different ways.}. The rescaled value is displayed in
the left panel of Fig.~\ref{fig:ideograms} as a vertical pink bar, for comparison with the ideogram. 

As an alternative, we present in the right panel of Fig.~\ref{fig:ideograms} a representation of the estimates
as a single Gaussian with the same normalization as the ideogram, and with its
central value and error chosen to reproduce the 95\% CL range of the 
ideogram as closely as possible:
\begin{equation}
{\rm Gaussian ~ representation:} \qquad \Sigma_{\pi N} \; = \; 46 \pm 11 \; {\rm MeV} \, .
\label{95percent}
\end{equation}
Although this functional form is far from perfect, we consider it a simple but fair representation
of current estimates of $\Sigma_{\pi N}$.

\begin{figure}[t!]
\centering
   \subfloat{\scalebox{0.7}{\includegraphics{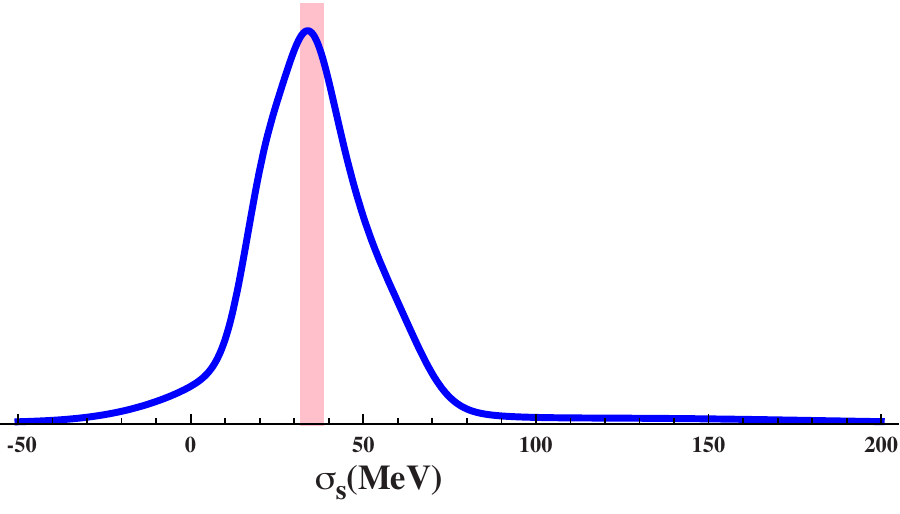}}}
\hspace{1cm}
    \subfloat{\scalebox{0.7}{\includegraphics{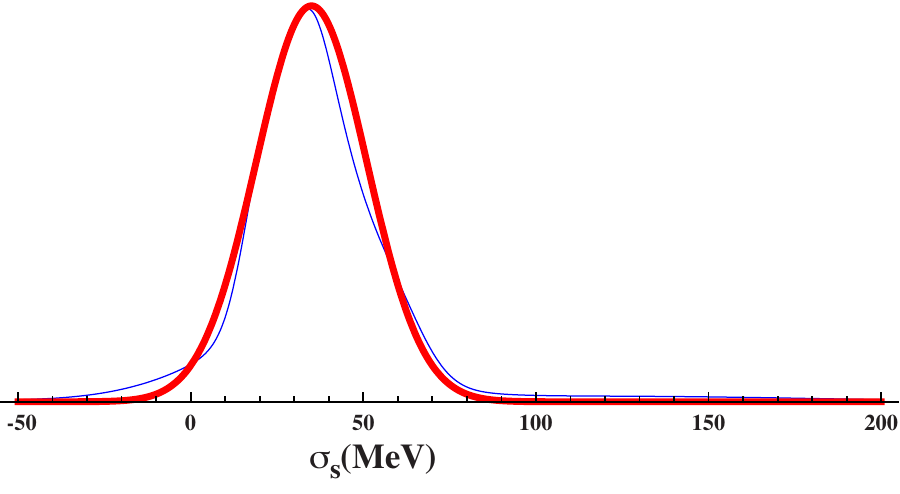}}}  
    \caption{\em Left panel: Ideogram combining all values of $\sigma_s$ .
    Right panel: Ideogram combining all values of $\sigma_s$
    compared with a Gaussian centred at $\sigma_s = 35$~MeV with width $\sigma = 16$~MeV.}
    \label{fig:ssideogram}
\end{figure}

Fig.~\ref{fig:ssideogram} shows the result of a similar exercise for the 23 determinations of $\sigma_s$ that we use.
In this case the ideogram has (barely) visible support out to very large values, but most of the numerical values
have support only for $\sigma_s < 100$~MeV. Following the same steps as used previously for $\Sigma_{\pi N}$, we find
\begin{eqnarray}
{\rm Naive:} \qquad \sigma_s & = & 35.2 \pm 2.6  \; {\rm MeV} \, , \nonumber \\
{\rm Rescaled:} \qquad \sigma_s & = & 35.2 \pm 3.1 \; {\rm MeV} \, , \nonumber \\
{\rm  Gaussian ~ representation:} \qquad \sigma_s & = & 35 \pm 16 \; {\rm MeV} \, .
\label{95percentss}
\end{eqnarray}
In this case the distribution obtained from the numerical estimates is again not symmetric, 
though it has a single-peak structure, and the representation
(\ref{95percentss}) may again be considered a simple but fair representation of
of current estimates of $\sigma_s$.

\subsection{Uncertainties in the Elastic Scattering Cross-Section}
\label{sec:densityMEs}

\begin{figure}[ht!]
\centering
   \subfloat{\scalebox{0.45}{\includegraphics{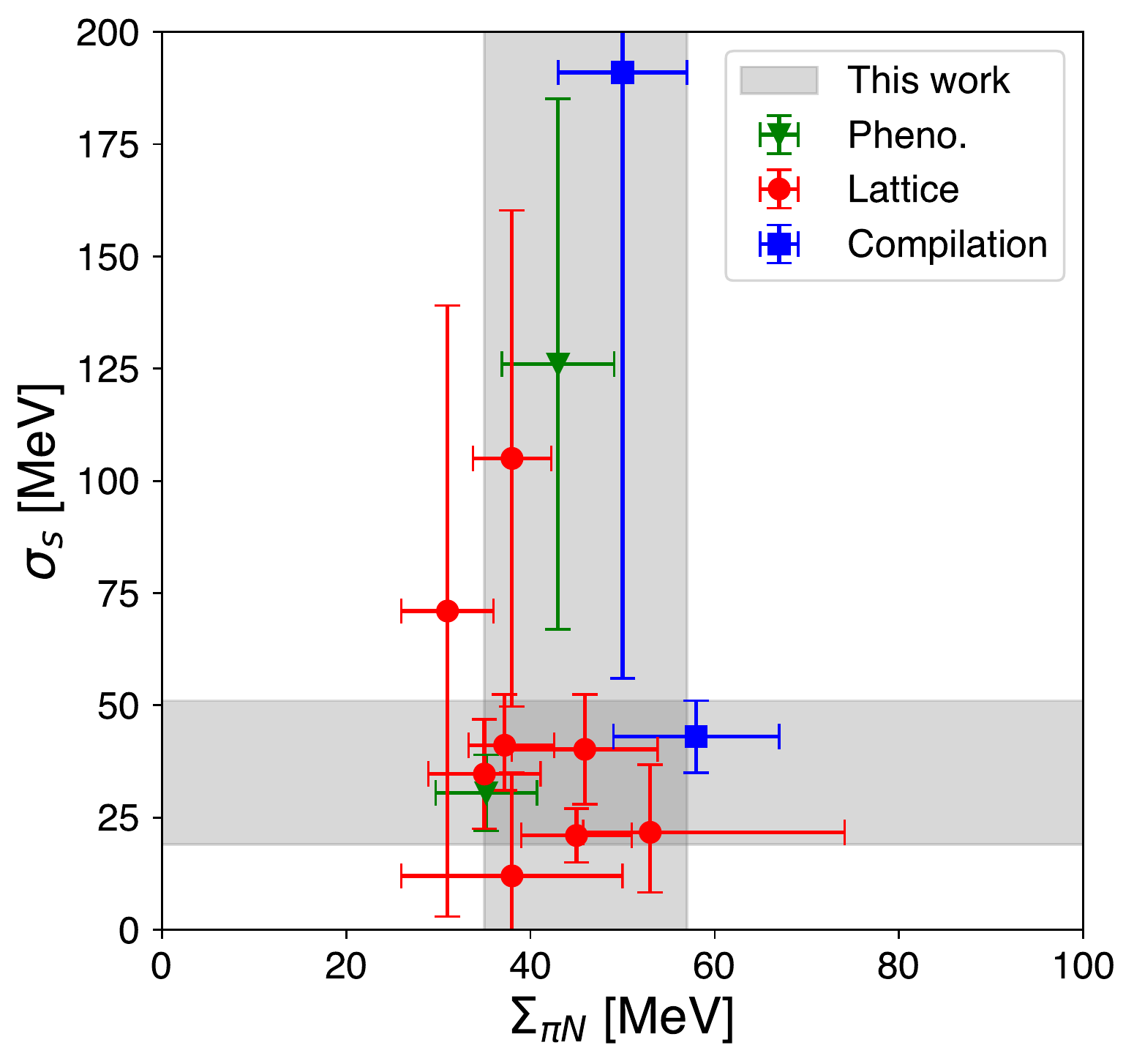}}}
   \hskip 0.25in
 \subfloat{\scalebox{0.55}{\includegraphics{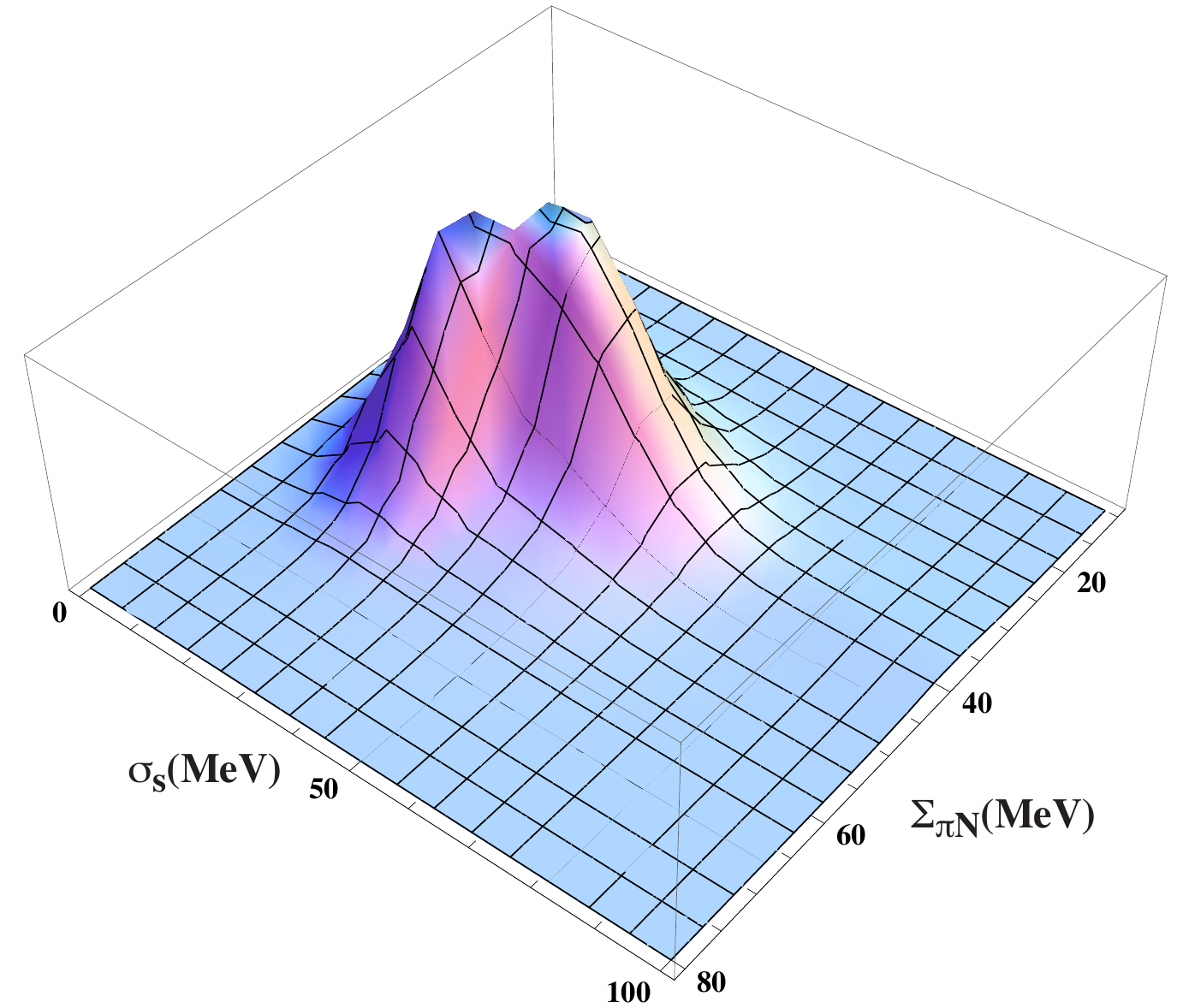}}} 
    \caption{\em Scatter plot of $(\Sigma_{\pi N}, \sigma_s)$, with grey
bands showing the ranges of $\Sigma_{\pi N}$ and $\sigma_s$ that we estimate on the basis of our new compilation. 
Right panel: Two-dimensional ideogram of $\Sigma_{\pi N}$ and $\sigma_s$.        }
    \label{fig:scatter}
\end{figure}

We now discuss the combination of the uncertainties in $\Sigma_{\pi N}$ and $\sigma_s$
in the calculation of $\sigma^p_{SI}$. A scatter plot of values of $(\Sigma_{\pi N}, \sigma_s)$ from references in which values of both quantities
are quoted (see Table~\ref{tab:values}) is shown in the left panel of Fig.~\ref{fig:scatter}. The horizontal and vertical grey
bands show the ranges of $\Sigma_{\pi N}$ and $\sigma_s$ that we estimate on the basis of our global analysis.
 The right panel of Fig.~\ref{fig:scatter} displays a two-dimensional 
joint ideogram of $\Sigma_{\pi N}$ and $\sigma_s$ based on the determinations in Fig.~\ref{fig:scatter}.
We have studied via a regression analysis whether these determinations exhibit any correlations between $\Sigma_{\pi N}$ and $\sigma_s$.
We find a slope in the $(\Sigma_{\pi N}, \sigma_s)$ plane of $0.49 \pm 1.08$, i.e., no significant correlation.
In the rest of this analysis we assume that there is no correlation between the uncertainties in $\Sigma_{\pi N}$ and $\sigma_s$.

\begin{figure}[ht!]
\centering
   \subfloat{\scalebox{0.35}{\includegraphics{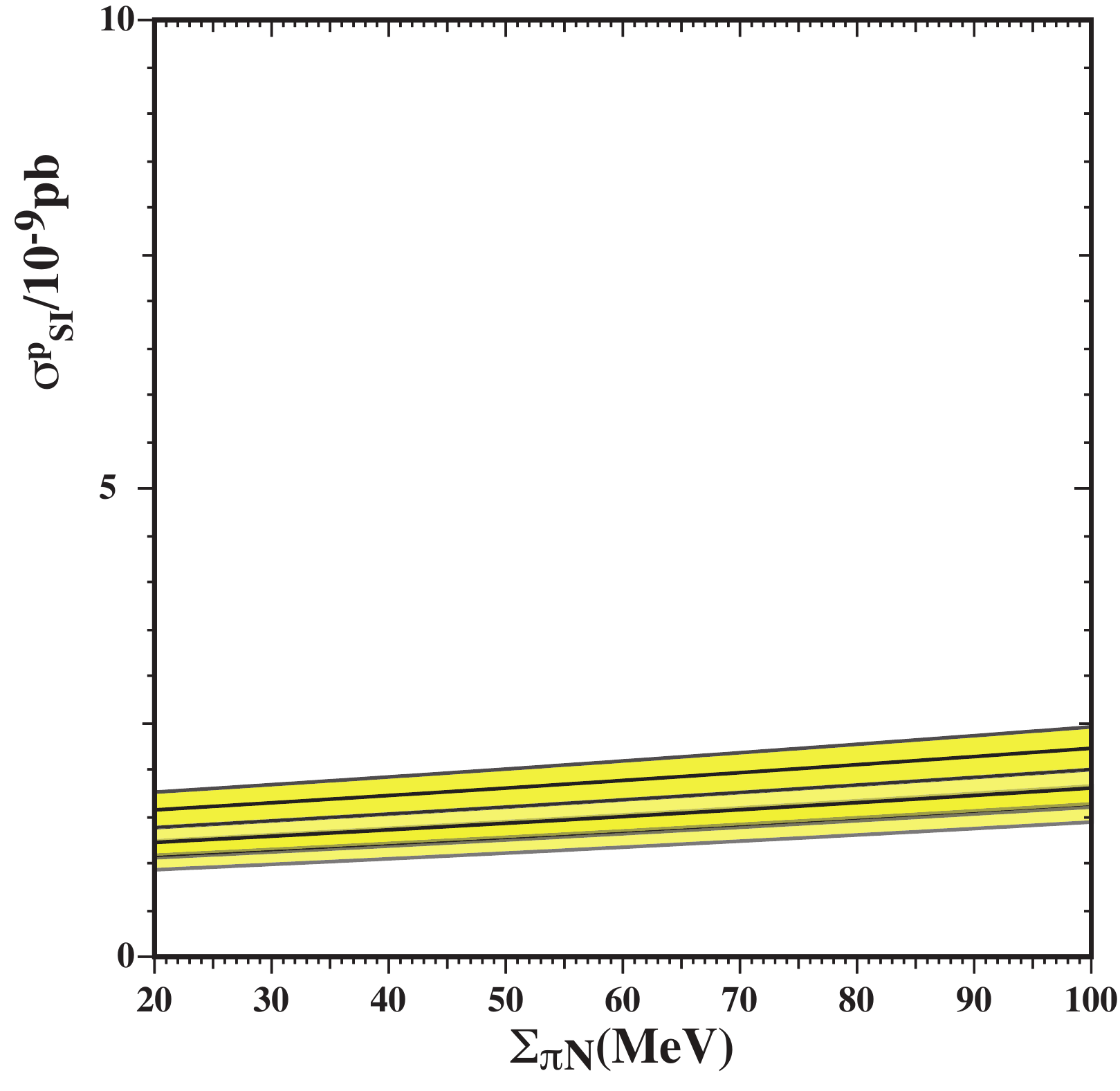}}}
\hskip .2in
    \subfloat{\scalebox{0.35}{\includegraphics{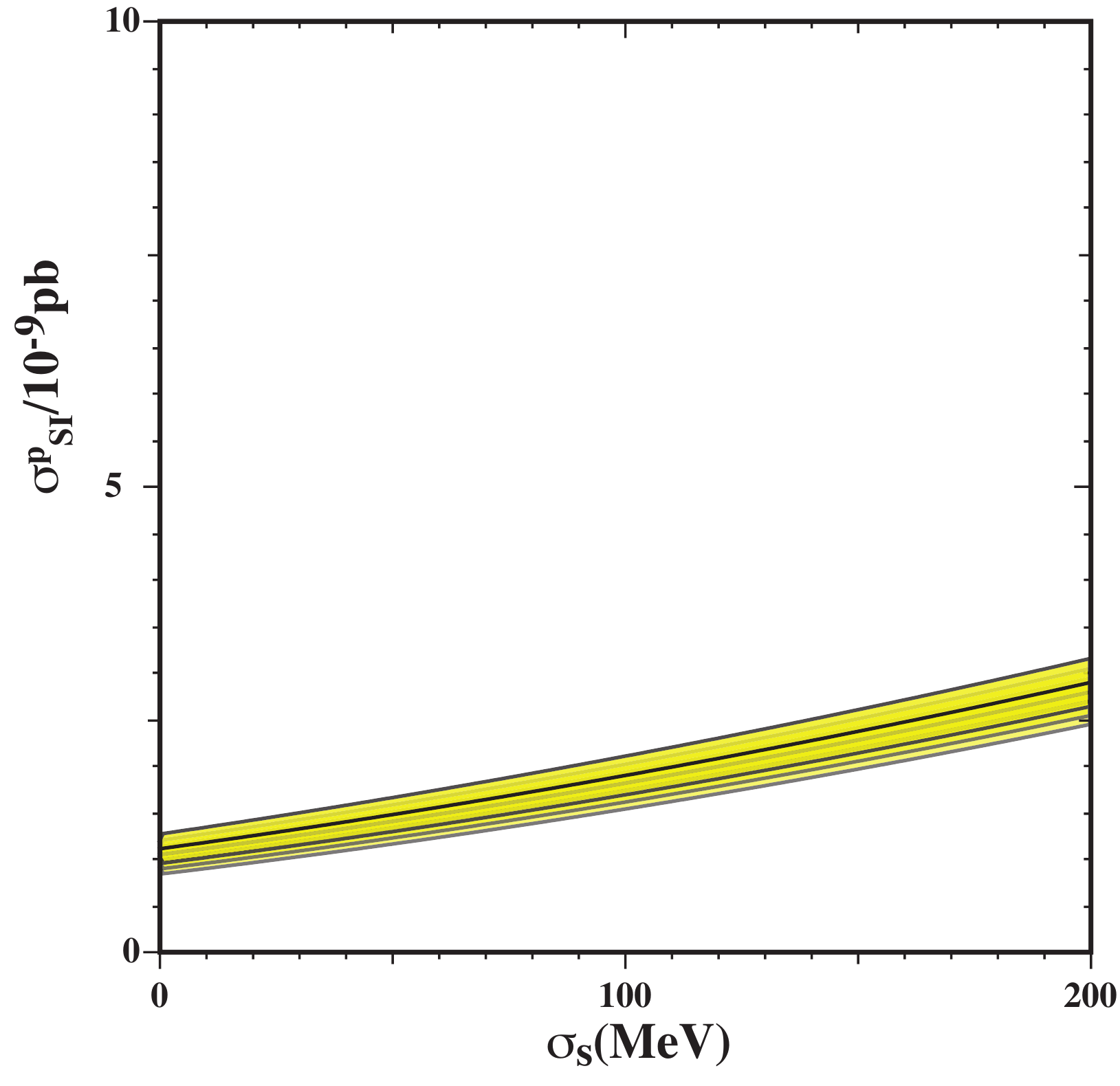}}}
    \caption{\em Left: $\sigma^p_{SI}$ vs $\Sigma_{\pi N}$ for $\sigma_s = 30, 50, 100$ MeV. 
    Right: $\sigma^p_{SI}$ vs $\sigma_s$ for  $\Sigma_{\pi N} =40, 50, 60$ MeV.
    The color bands show the calculated 1 $\sigma$ uncertainty in the elastic cross section. 
        }
    \label{fig:sigmapiNsigmas}
\end{figure}

The analogue of Fig. \ref{fig:sigma0sigmas} in terms of $\Sigma_{\pi N}$ and $\sigma_s$
is shown in Fig. \ref{fig:sigmapiNsigmas}, where the values of $\sigma^p_{SI}$ are again calculated using the three-flavour expression (\ref{f3}).
In the left panel, we see that for fixed $\sigma_s$, there is no longer the large dependence of the cross section 
on $\Sigma_{\pi N}$ that was seen in Fig. \ref{fig:sigma0sigmas}. The three bands (which overlap) correspond to 
$\sigma_s = 30, 50$ and 100 MeV.  In the right panel, there are three bands corresponding to fixed 
values of $\Sigma_{\pi N} = 40, 50$ and 60 MeV that lie almost on top of each other, and one sees quite 
clearly the dependence of the cross section on $\sigma_s$.  Note that the thickness of the bands here
are significantly narrower than those in Fig. \ref{fig:sigma0sigmas}. This is mainly due to the smaller uncertainty in 
$\sigma_s$ when it is taken directly from Eq. (\ref{95percentss}) rather than derived indirectly using estimates of $\sigma_0$.
Using $\sigma_s$ as an input into calculations of $\Sigma_{\pi N}$ is therefore preferred.

We now show how the uncertainties in $\Sigma_{\pi N}$ and $\sigma_s$
discussed above can be propagated into the uncertainty in the elastic scattering cross section.
As we have discussed, the distributions for the hadronic matrix elements are not
Gaussian, but we have provided Gaussian approximations to those distributions in 
(\ref{95percent}) and (\ref{95percentss}), which we propagate to the errors in
the cross section. We describe our procedure for scattering on protons only,
the neutron case being simply related by an isospin transformation.

The uncertainty in the elastic cross section on a proton is simply
\beq
\sigma_{\sigma^p_{SI}} = 2 \sigma^p_{SI} \frac{\sigma_{f_p}}{f_p} \, ,
\eeq
where
\beq
\sigma_{f_p} = m_p \left(\sum_{q=u,d,s} \sigma_{f^p_{T_q}}^2 \left[(\alpha_{3_q}/m_q)               
    - \frac{2}{27}  (\sum_{q=c,b,t} \alpha_{3_q}/m_q )\right]^2 \right)^{1/2} \, .
\label{sigmaf}    
\eeq
Since we are using $\sigma_s$ and its uncertainty directly, calculating the uncertainty in $f^p_{T_s} = \sigma_s/m_p$ is
straightforward:
\beq
\sigma_{f^p_{T_s}} = f^p_{T_s} \frac{\sigma_{\sigma_s}}{\sigma_s} \, .
\eeq
In the cases of the other light quarks,
the expressions for $f^p_{T_q}$ are more complicated:
\begin{eqnarray}
m_p f^p_{T_u} & =  & \frac{2 \Sigma_{\pi N}}{(1+\frac{m_d}{m_u})(1+\frac{B_d}{B_u})}  =  \frac{2 m_u}{m_u+m_d} \left[
\frac{z}{1+z}\Sigma_{\pi N} + \frac{m_u+m_d}{2 m_s} \frac{1-z}{1+z} \sigma_s \right] \nonumber \\
m_p f^p_{T_d} &  =  & \frac{2 \Sigma_{\pi N}}{(1+\frac{m_u}{m_d})(1+\frac{B_u}{B_d})}  =    \frac{2 m_d}{m_u+m_d} \left[
\frac{1}{1+z}\Sigma_{\pi N} - \frac{m_u+m_d}{2 m_s} \frac{1-z}{1+z} \sigma_s \right]  \, ,
\end{eqnarray}
where the right-hand sides of the equations allow us to compute the $f^p_{T_q}$ directly from our inputs.
To obtain the uncertainties in $f^p_{T_{u,d}}$, we propagate the
uncertainties in $\Sigma_{\pi N}$, the light quark mass ratio $m_u/m_d = 0.46 \pm 0.05$ (\ref{PDGratios}),
and in the ratio $B_d/B_u$ given in Eq. (\ref{eqn:BdBu}). This depends on the uncertainty in $y$,
and hence depends ultimately on the uncertainties in
$m_s/(m_u+m_d) = 13.75 \pm 0.15$ and $\sigma_s$. 
The expression (\ref{sigmaf}) takes into account the correlation in the uncertainties between the light and heavy
quark contributiions. 

We have verified in the benchmark model assumed 
that the uncertainties in $m_u/m_d$ and in $m_s/m_d$ given in (\ref{udsmasses}) contribute very
small uncertainties to $\sigma^p_{SI}$, a few per mille and below one per mille respectively. 
The uncertainty due to $B^p_d/B^p_u$ is also small, at the $\pm 2$\% level for $1 < z < 2$~\footnote{Over this range of $z$,
$\sigma^p_{SI}/\sigma^n_{SI}$ varies between 1.00 and 0.94.}. We note that our benchmark point is taken from
a supersymmetric theory and the scattering of the dark matter candidate in this model on a proton is dominated by
the heavy quark content. It is quite possible that other dark matter candidates are more sensitive to the scattering
off of light quarks and in that case, the uncertainty due to $B^p_d/B^p_u$  and $z$ is more important.

\begin{figure}[t!]
\centering
    \subfloat{\scalebox{0.55}{\includegraphics{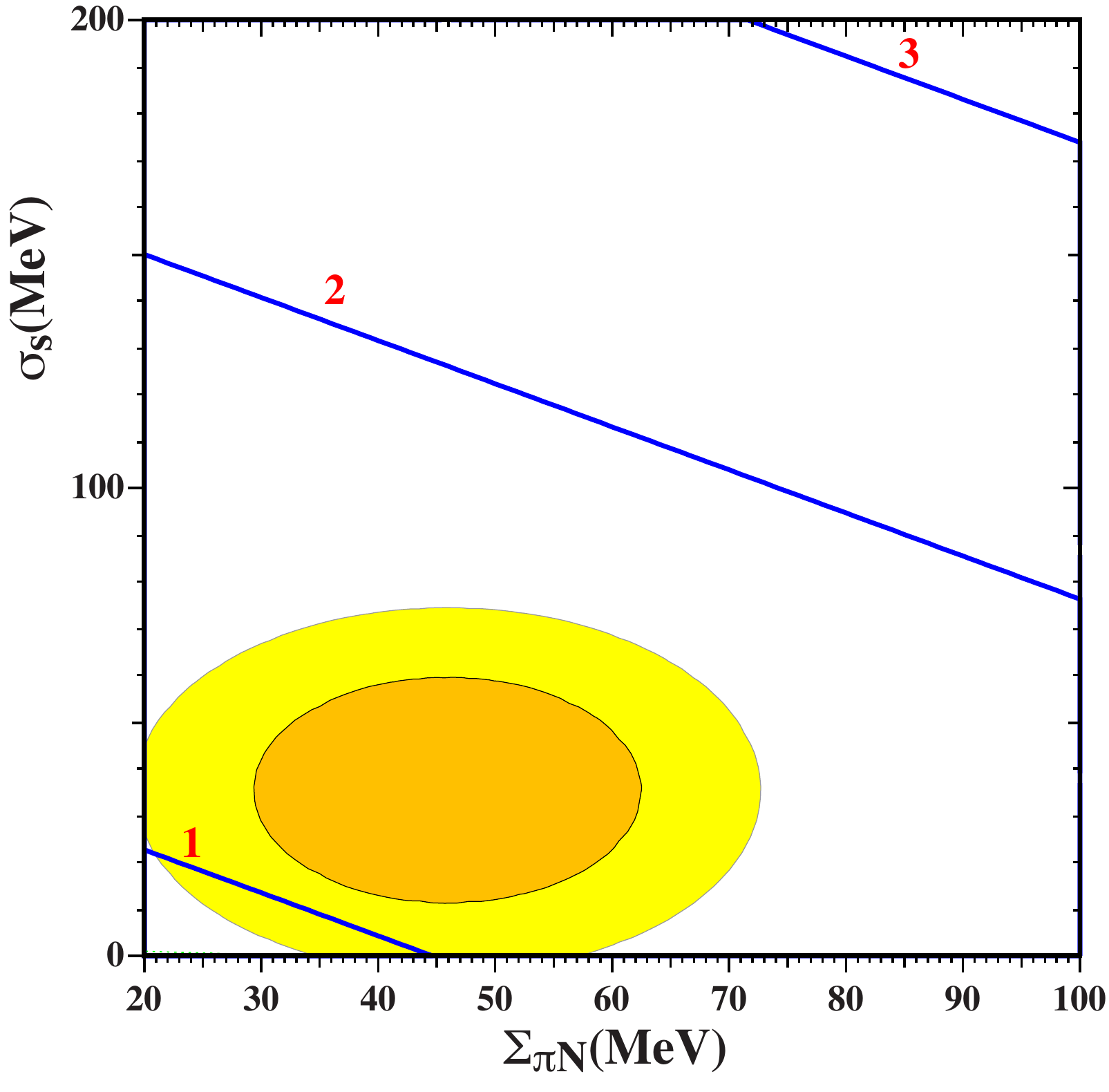}}}  
    \caption{\em 
    Contours of $\sigma^p_{SI}$ (in units of $10^{-9}$~pb) in the $(\Sigma_{\pi N}, \sigma_s)$ plane, with the
    two-dimensional 68\% and 95\% CL regions (darker and lighter shading) given by our Gaussian fits (\protect\ref{95percent}, \protect\ref{95percentss})
    to $\Sigma_{\pi N}$ and $\sigma_s$.}
    \label{fig:plane}
\end{figure}

We display in Fig.~\ref{fig:plane} contours of $\sigma^p_{SI}$ (in units of $10^{-9}$ pb) in the $(\Sigma_{\pi N}, \sigma_s)$ plane
calculated using the three-flavour expression (\ref{f3}), 
together with the two-dimensional 68\% and 95\% CL regions ($\Delta \chi^2 < 2.3$ and $5.99$,
respectively) given by our Gaussian fits (\ref{95percent}, \ref{95percentss})
to $\Sigma_{\pi N}$ and $\sigma_s$, assuming that there is no correlation, as discussed above.

Using our values for $\Sigma_{\pi N}$ (\ref{95percent}) and $\sigma_s$ (\ref{95percentss})
that are also given in the last line of Table \ref{tab:values}, we find
$\sigma^p_{SI} = (1.25 \pm 0.13) \times 10^{-9}$ pb when we use the three-flavour expression (\ref{f3}) 
for our CMSSM benchmark point. The decrease in the cross section (by a factor of 2) relative to what we would have calculated using
the values of $\Sigma_{\pi N}$ and $\sigma_0$ used in~\cite{mc12}
is due largely to the effective reduction in $\sigma_s$. Moreover, the uncertainty in the cross section is
a factor of 10 smaller. This reduction can be traced to using $\sigma_s$ (and its uncertainty) directly from
the recent calculations - as we recommend - rather than using the value inferred from (\ref{sigmas}) and the older values of $\Sigma_{\pi N}$ and $\sigma_0$.

\subsection{Dependence on heavy quark matrix elements}
\label{sec:HQ}

In this Section, we explore the sensitivity of $\sigma^p_{SI}$ to the heavy quark matrix elements,
using first the four-quark version (\ref{f4}) of the cross-section formula, and then the full six-flavour version.

\begin{figure}[ht!]
\centering
   \subfloat{\scalebox{0.5}{\includegraphics{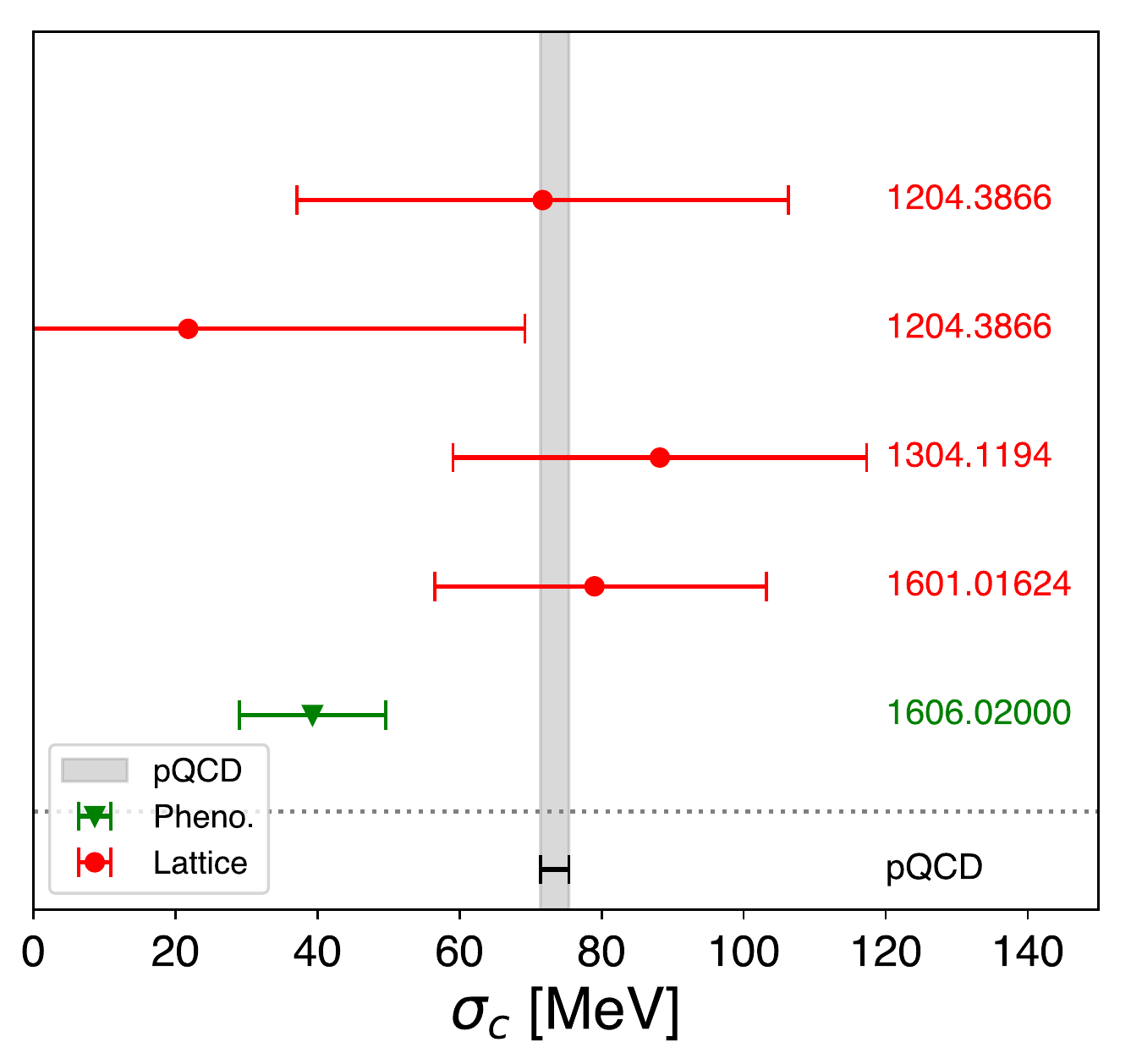}}}
    \caption{\em Calculations of $\sigma_c$ (in MeV units). 
    Those made using the lattice are indicated in red, and one relying on
    phenomenological inputs is indicated in green.
    We also show the ${\cal O}(\alpha_s^3)$ perturbative calculation (\protect\ref{sigmac}) (bottom line and vertical grey band).
        }
    \label{fig:Natsumic}
\end{figure}

\begin{figure}[ht!]
\centering
    \subfloat{\scalebox{0.65}{\includegraphics{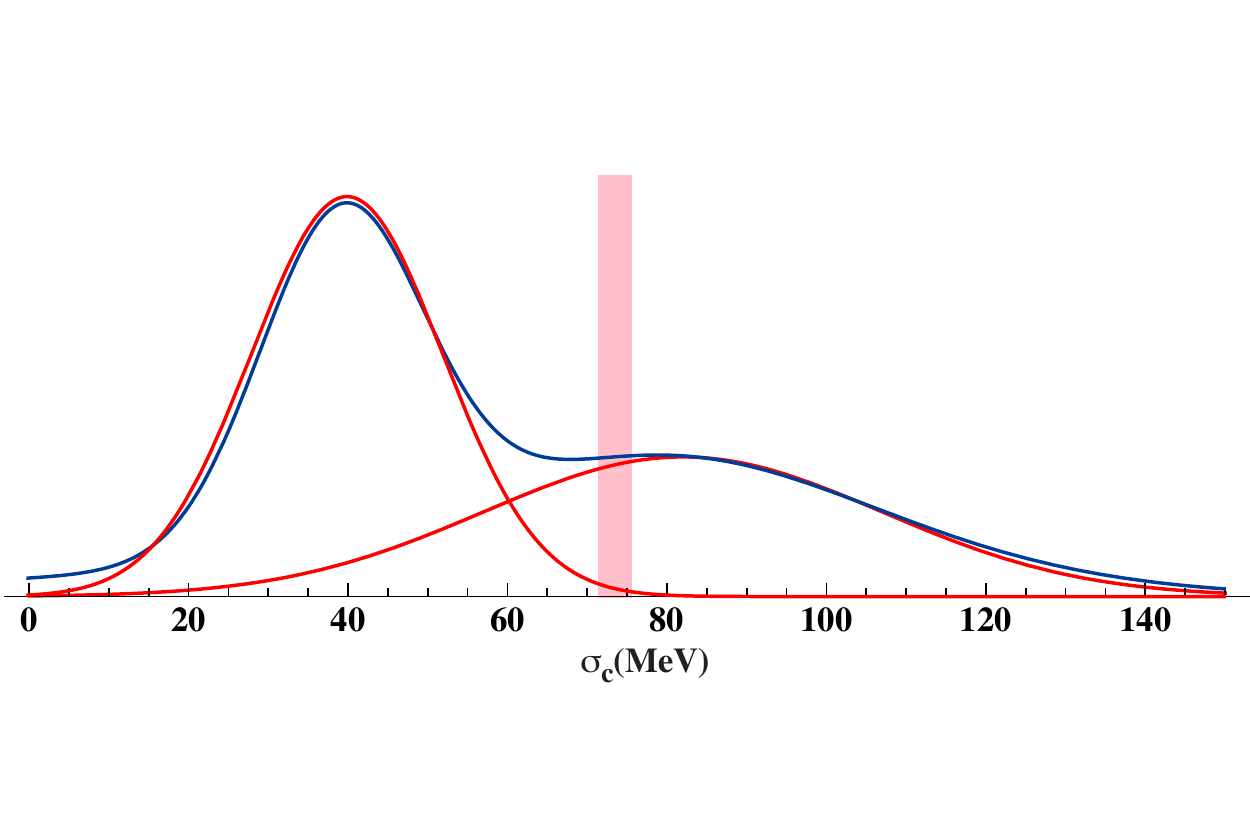}}}
\vspace{-1cm}
    \caption{\em Ideogram compiling lattice and phenomenological calculations of $\sigma_c$
    and our representation as a sum of two Gaussians (\protect\ref{95percentcc}).
    Also shown is a vertical pink band from our evaluation of the ${\cal O}(\alpha_s^3)$ perturbative 
    QCD calculation of~\protect\cite{vecchi}, shown in (\protect\ref{sigmac}).}
    \label{fig:sigmac}
\end{figure}

There have been several calculations of the charm quark contribution to the proton mass,
$\sigma_c \equiv m_c \langle N | {\bar c} c | N \rangle$, most using lattice techniques, as shown
in Fig.~\ref{fig:Natsumic} and Table~\ref{tab:HQvalues}. There have also been some phenomenological
estimates of $\sigma_c$, as also shown there. We do not include in our global fit the last
three estimates, which depend on particular hypotheses concerning the fraction of the proton momentum in the
infinite-momentum frame that is carried by ${\bar c} c$ pairs~\cite{Hobbs:2017fom}.
In the absence of a clear criterion for choosing between
these hypotheses, we do not use them. An ideogram of the estimates of $\sigma_c$
that we retain is shown in Fig.~\ref{fig:sigmac}.
Following the same steps as used previously for $\Sigma_{\pi N}$ and $\sigma_s$, we find
\begin{eqnarray}
{\rm Naive:} \qquad \sigma_c & = & 50.2 \pm 8.5  \; {\rm MeV} \, , \nonumber \\
{\rm Rescaled:} \qquad \sigma_c & = & 50.2 \pm 9.6 \; {\rm MeV} \, . \nonumber
\end{eqnarray}
However, it is clear that these are exceedingly poor representations of the ideogram
shown in Fig.~\ref{fig:sigmac}, which is highly asymmetric. We choose to represent
this by summing a pair of Gaussians $G$, with arbitrary normalizations and the following central values and errors:
the maximum:
\begin{equation}
{\rm Gaussian ~ representation:} \qquad \sigma_c \; = \; G ( 40~{\rm MeV}, 12~{\rm MeV}) +  G ( 82~{\rm MeV}, 25~{\rm MeV})\, .
\label{95percentcc}
\end{equation}
This representation is also shown in Fig.~\ref{fig:sigmac}, and gives a very good representation of the estimates in Table~\ref{tab:HQvalues}.

\begin{table}[ht!]
\caption{Estimates of $\sigma_{c}$. }
\label{tab:HQvalues}
\vspace{-0.5cm}
\begin{center}
\scriptsize
\begin{tabular}{|c||c|c||c|}
\hline
Reference & $\sigma_c$ & Uncertainties &  Method \\
\hline
\cite{Freeman:2012ry} & 71.7 & 34.6 & Lattice \\
	&21.8 & 47.4 & Lattice \\
	\hline
\cite{Gong:2013vja} & 88.2 & 29.1 & Lattice \\
\hline
\cite{Abdel-Rehim:2016won} & 79 & $21 ^{+12}_{-8}$ & Lattice \\
\hline
\cite{Duan:2016rkr} & 39.3 & 10.3 & Phenomenology \\
\hline
\hline
\cite{Hobbs:2017fom} & 4.3 & 4.4 & Phenomenology\\
	& 12.5 & 13 & Phenomenology \\
	& 32.3 & 33.6 & Phenomenology \\
\hline
\end{tabular}
\end{center}
\caption*{\it Calculations of $\sigma_{c}$ (in MeV units). Where two
errors are quoted, the first is statistical and the second systematic. As indicated, most of the determinations
are based on lattice calculations. The last group of three phenomenological estimates are not included
in our global fit.
}
\end{table}

Using the four-quark expression (\ref{f4}) and the first Gaussian for $\sigma_c$ in (\ref{95percentcc}),
we find $\sigma^p_{SI} = (1.07 \pm 0.15) \times 10^{-9}$~pb, whereas the second Gaussian in (\ref{95percentcc})
yields $\sigma^p_{SI} = (1.40 \pm 0.25) \times 10^{-9}$~pb, reflecting its larger central value and error. 
We note that, in the computation of these uncertainties, Eq. (\ref{sigmaf}) must be modified in a way similar to Eq. (\ref{f4}), namely
the first sum is over the four quarks $u,d,s,c$, $2/27 \to 2/25$, and the second sum is over two quarks $b,t$. 

\begin{figure}[t!]
\centering
    \subfloat{\scalebox{0.35}{\includegraphics{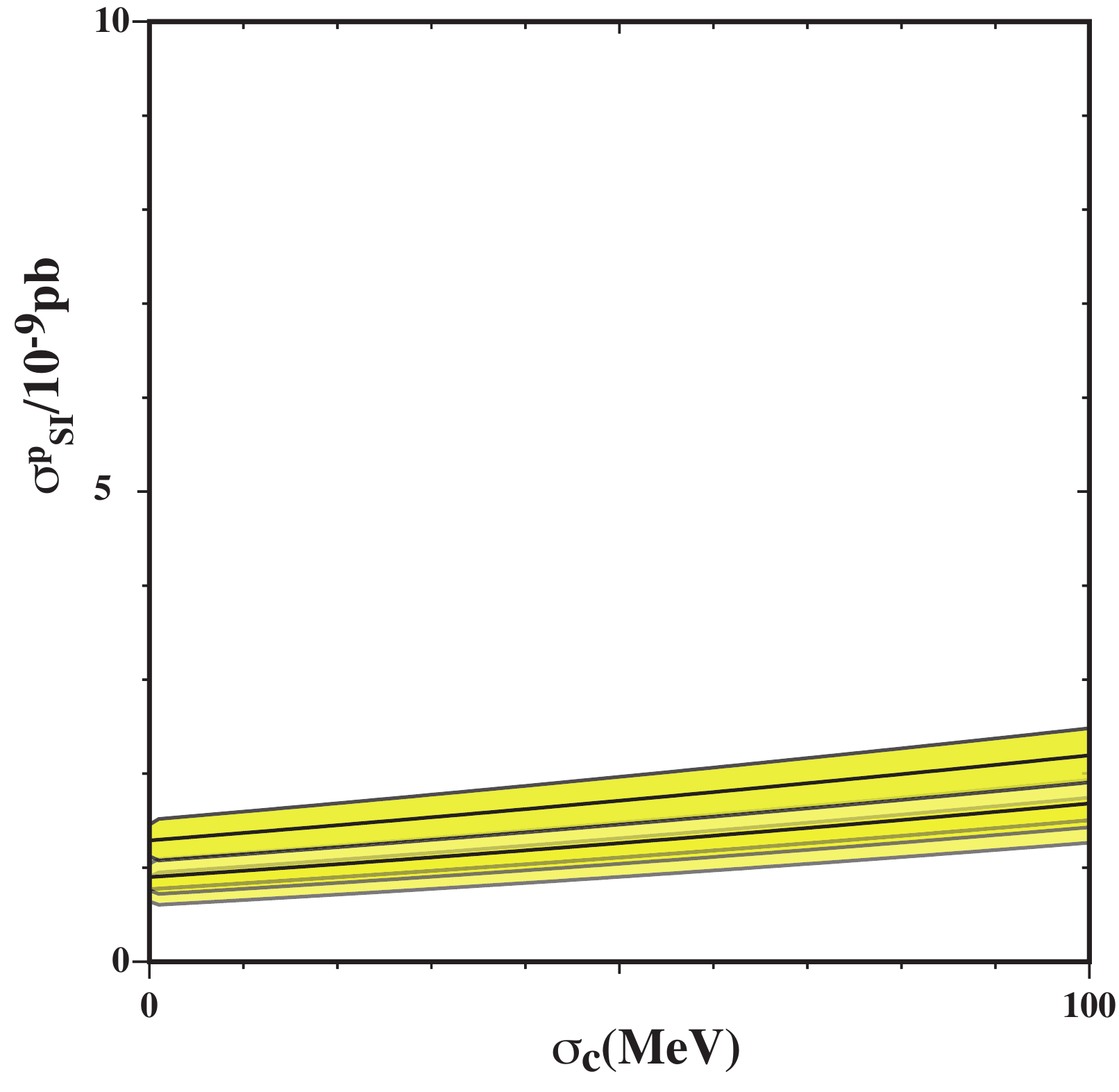}}}
  \hskip .2in
     \subfloat{\scalebox{0.35}{\includegraphics{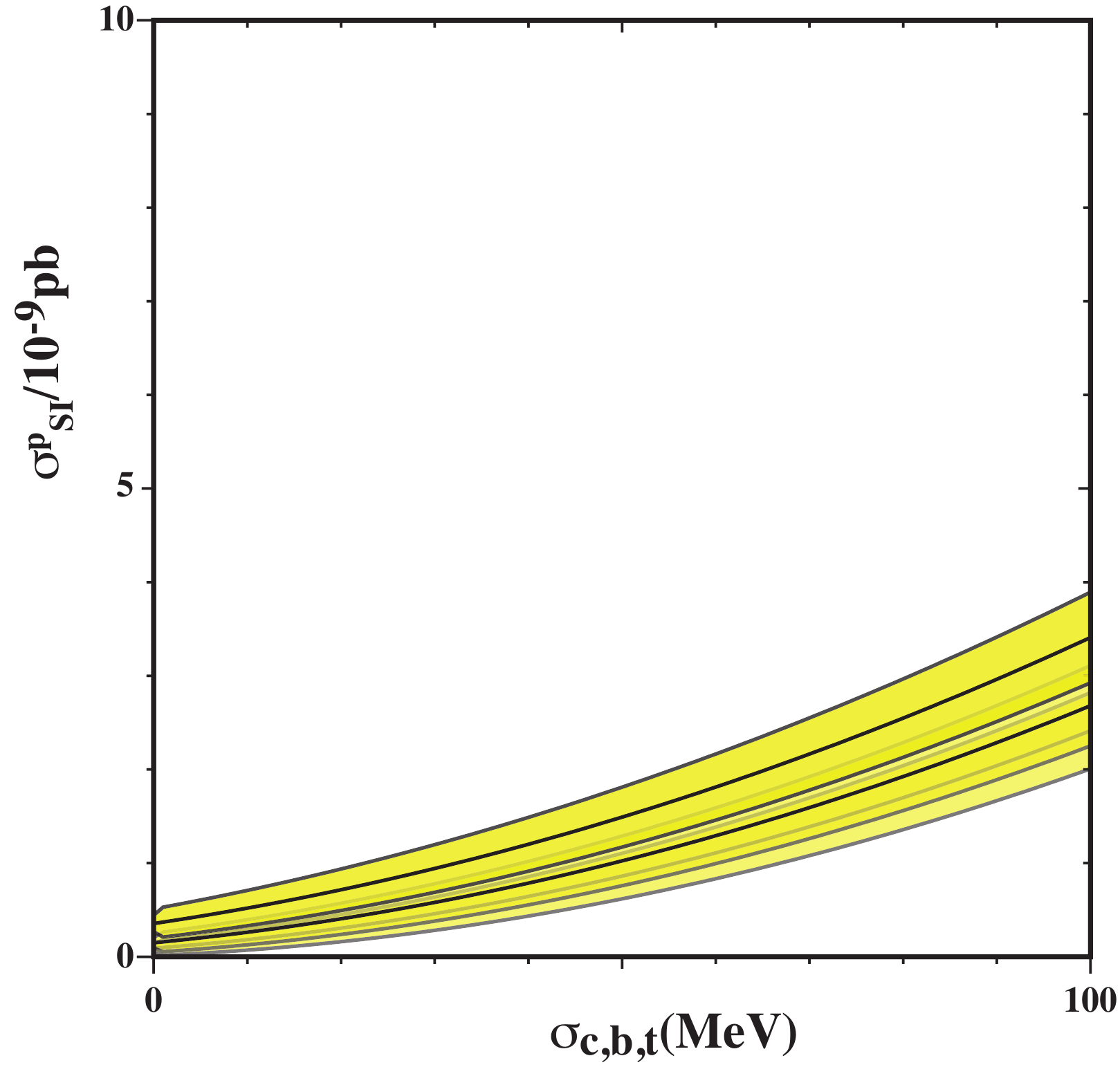}}}
    \caption{\em Left: $\sigma^p_{SI}$ vs $\sigma_c$ for fixed $\Sigma_{\pi N} = 46$~MeV and $\sigma_s = 30, 50, 100$~MeV. 
    Right: $\sigma^p_{SI}$ vs $\sigma_c = \sigma_b = \sigma_t$ for fixed $\Sigma_{\pi N} = 46$~MeV and $\sigma_s = 30, 50, 100$~MeV. 
     }
    \label{fig:sigmacbt}
\end{figure}

Alternatively, one could adopt the value for $\sigma_c$ taken from a perturbative QCD calculation using the one-loop
contribution associated with the gluon contribution \cite{SVZ}
so that~\footnote{Here and in the rest of this Section we give results for $\sigma_{c,b,t}$ in the proton. Becasue of isospin violation,
perturbative calculations of the central values in the neutron yield slightly different results, but  these are indistinguishable within the uncertainties. 
We report results for the $f^{p,n}_{T_{c,b,t}}$ separately in Table~\ref{tab:fTvalues} below.}:
\beq
\sigma_c = \frac{2}{27}M_N f^p_{T_G} = 69.5 f^p_{T_G} \, .
\eeq
Using our estimates (\ref{95percent}, \ref{95percentss}) of $\Sigma_{\pi N}$
and $\sigma_s$, we find $f^p_{T_G} = 0.917 \pm 0.019$ and hence
\beq
\sigma_c = (63.7 \pm 1.3) \, {\rm MeV} \, ,
\label{sigmacp}
\eeq
which lies between are two Gaussian estimates of $\sigma_c$. 
Not surprisingly, the cross section calculated this way, $\sigma^p_{SI} = (1.25 \pm 0.14) \times 10^{-9}$~pb,
is almost identical to our 3-quark calculation using $\Sigma_{\pi N}$ and $\sigma_s$.
In fact, this approach does not distinguish between the contributions of the heavy 
quarks and would imply $\sigma_b = \sigma_t = \sigma_c$ at this order in perturbation theory.
Using Eq. (\ref{sigmacp}) for all 3 heavy quarks gives $\sigma^p_{SI} = (1.24 \pm 0.17) \times 10^{-9}$~pb.
Fig.~\ref{fig:sigmacbt} displays the sensitivity of $\sigma^p_{SI}$ to the assumed value of $\sigma_c$
in the four-quark formula (left panel) and $\sigma_c = \sigma_b = \sigma_t$ in the six-quark formula (right panel). 
In both panels we have set $\Sigma_{\pi N} = 46$ MeV, and the 3 bands correspond again to 
$\sigma_s = 30, 50$ and 100 MeV.

One can go beyond the above 1-loop calculation and improve the perturbative QCD calculation for $\sigma_c$ by going to 
 ${\cal O}(\alpha_s^3)$~\cite{Kryjevski:2003mh,vecchi,hill}. Using Eq. (2.13) from \cite{vecchi} and
 $f^p_{T_G} = 0.917 \pm 0.019$, we find
\begin{equation}
\sigma_c \; = \; \frac{2}{27} \left(-0.3 + 1.48 f^p_{T_G} \right) m_p  \; = \; 73.4 \pm 1.9 \; {\rm MeV} \, .
\label{sigmac}
\end{equation}
Also shown in Fig.~\ref{fig:sigmac} is a vertical pink band
corresponding to this evaluation of the ${\cal O}(\alpha_s^3)$ perturbative QCD calculation~\cite{vecchi}.
It has been argued (see~\cite{Brodsky} for a review) that there may be non-perturbatively-generated intrinsic charm in the nucleon,
in which case the perturbative calculation leading to (\ref{sigmac})
would be inapplicable. 
Another potential source of difference is caused by higher-dimensional
operators that are induced when the charmed quark is integrated out \cite{vecchi}, which
are suppressed only by the charmed quark mass and thus may give a
significant contribution to the nucleon mass. 
The difference
between (\ref{95percentcc}) and (\ref{sigmac}) may serve as a measure of the uncertainty associated
with these possibilities.

Similar ${\cal O}(\alpha_s^3)$ perturbative QCD calculations for the $b$ and $t$ quarks are expected to be more reliable, and yield~\cite{vecchi}:
\begin{eqnarray}
\sigma_b & = & \frac{2}{27} \left(-0.16 + 1.23 f^N_{T_G} \right) M_N \, , \\
\sigma_t & = & \frac{2}{27} \left(-0.05 + 1.07 f^N_{T_G} \right) M_N \, ,
\label{bt}
\end{eqnarray}
which, in combination with our estimates (\ref{95percent}, \ref{95percentss}) of $\Sigma_{\pi N}$ and $\sigma_s$, and $f^p_{T_G}$ yield 
\begin{equation}
\sigma_b \; = \; 67.3 \pm 1.6 \; {\rm MeV} \; , \; \sigma_t \; = \; 64.7 \pm 1.4 \; {\rm MeV} \, ,
\label{btnumbers}
\end{equation}
for the proton.
Using the ${\cal O}(\alpha_s^3)$ perturbative calculations of $\sigma_c$ (\ref{sigmac}) and of 
$\sigma_{b,t}$ (\ref{btnumbers}) in a full six-flavour calculation, we find
$\sigma^p_{SI} = (1.38 \pm 0.17) \times 10^{-9}$~pb. The origin of the increase from the three-flavour 
approximation to the full six-flavour calculation is primarily the fact that the more detailed perturbative
treatment of the heavy quarks increases their contributions, particularly that of the charmed quark, by ${\cal O}(15)$\%.

We consider the full six-flavour calculation using the estimates (\ref{95percent}, \ref{95percentss}, \ref{sigmac}) and (\ref{btnumbers})
to be the best approximation to the spin-independent WIMP scattering cross section currently available.

\section{Spin-Dependent WIMP-Nucleon Scattering}
\label{sec:SD}

In the case of the cross section $\sigma_{SD}$ for spin-dependent WIMP-nucleon scattering, the relevant matrix elements
$\langle N |{\bar q}_i \gamma_\mu \gamma_5 q_i | N \rangle$ are related to the
corresponding quark contributions to the nucleon spin $\Delta q_i$. The combination
$\Delta u - \Delta d = g_A = 1.27$, the axial-current matrix element in neutron $\beta$-decay,
which is known quite precisely. We estimate the combination $\Delta u + \Delta d - 2 \Delta s = 0.59$
using other octet baryon weak decay matrix elements and SU(3) symmetry. A third combination of 
the light-quark $\Delta q_i$ can be determined from parity-violating asymmetries in
polarized deep-inelastic electron- and muon-nucleon scattering~\cite{EFR}, which indicate a small
but non-zero negative value of $\Delta s = - 0.09 \pm 0.03$ when combined with the above-mentioned
estimated of $\Delta u - \Delta d$ and $\Delta u + \Delta d - 2 \Delta s$. Measurements of hadron
production asymmetries in polarized deep-inelastic scattering do not support a non-zero
value of $\Delta s$. Nevertheless, this confusion in the
estimates of the $\Delta q_i$ generates only moderate uncertainty in the cross section
for spin-dependent WIMP-nucleon scattering, $\sigma_{SD}$.

For the CMSSM focus-point benchmark point introduced above, we find that the value $\Delta s = - 0.09 \pm 0.03$
indicated by the parity-violating asymmetries in the total polarized deep-inelastic cross sections
leads to $\sigma^p_{SD} = (9.4 \pm 0.8) \times 10^{-7}$~pb, whereas the choice $\Delta s = 0 \pm 0.03$ would yield
$\sigma^p_{SD} = (8.2 \pm 0.7) \times 10^{-7}$~pb. The uncertainty in the spin-dependent cross section is
largely determined by the uncertainty in $\Delta_s$, and ignoring the uncertainty in $\Delta_s$ would reduce the uncertainty in 
$\sigma^p_{SD}$ to $\pm 0.2$.  The corresponding cross-section for scattering off neutrons is
 $\sigma^p_{SD} = (7.1 \pm 0.7) \times 10^{-7}$~pb for  $\Delta s = - 0.09$. When $\Delta s = 0$, there is virtually
 no difference between the cross sections for scattering on protons and neutrons.
We conclude that the uncertainties in
spin-dependent WIMP-nucleon scattering are comparable to the current uncertainties in 
spin-independent WIMP-nucleon scattering that have been the main focus of this paper.

\section{Conclusions}
\label{sec:conx}

We have re-analyzed in this paper ingredients in the calculation of the cross section for the spin-independent scattering
of a massive WIMP on a nucleon. Based on available recent calculations using lattice and other techniques,
we have used the prescription of the PDG to discuss the uncertainties in the quark
scalar densities $\langle N | {\bar q} q | N \rangle$. We find a central value for the combination $\Sigma_{\pi N}$ of 
$u$ and $d$ densities that is somewhat smaller than found in previous compilations~\cite{eosv,mc12,gambit}, 
though with a larger uncertainty: $\Sigma_{\pi N} = 46 \pm 11$~MeV. All determinations are compatible within the stated errors. 
We also find $\sigma_s = m_s \langle N | {\bar s} s | N \rangle = 35 \pm 16$~MeV,
which is again smaller than suggested in previous compilations, with an uncertainty that is significantly smaller than in~\cite{eosv,mc12}
but somewhat larger than in~\cite{gambit}. We find (for the benchmark supersymmetric model we have studied)
that the uncertainty in $\sigma_s$ is the largest single source of uncertainty in 
$\sigma_{SI}$ when it is calculated using the leading-order three-flavour approximation (\ref{f3}) for the spin-independent
scattering matrix element~\footnote{
As we noted above, the uncertainty due to $B_d^p/B_u^p$ and $z$ may be
more important in models where the spin-independent scattering occurs
primarily off $u$ and $d$ quarks. 
}. The corresponding values of the $f^N_{T_{u,d,s,G}}$ for scattering on protons and neutrons
obtained assuming $z = 1.49$ are shown in the first four columns of Table~\ref{tab:fTvalues}, 
where we include the uncertainties due to the $u, d, s$ mass ratios.
Note that although the spin-independent cross section is not particularly sensitive
to $z$, the values of $f^N_{T_{u,d}}$ do depend on $z$. However, for $1 < z < 2$, their central values 
vary within the 1$-\sigma$ ranges quoted in Table~\ref{tab:fTvalues}. Specifically, for $z = 1 (2)$, 
we find $f^p_{T_u} = 0.015 \pm 0.004$ $(0.020 \pm 0.005)$
and $f^p_{T_d} = 0.034 \pm 0.008$ $(0.023 \pm 0.006)$.

\begin{table}[ht!]
\caption{Values of the $f^{N}_{T_{q,G}}$. }
\label{tab:fTvalues}
\vspace{-0.5cm}
\begin{center}
\begin{tabular}{|c||c|c|c|c||c|c|c|}
\hline
Nucleon & $f^N_{T_u}$ & $f^N_{T_d}$ & $f^N_{T_s}$ & $f^N_{T_G}$ & $f^N_{T_c}$ & $f^N_{T_b}$ & $f^N_{T_t}$\\
\hline
Proton & 0.018(5) & 0.027(7) & 0.037(17) & 0.917(19) &0.078(2) & 0.072(2) & 0.069(1) \\
Neutron & 0.013(3) & 0.040(10) & 0.037(17) & 0.910(20) & 0.078(2) & 0.071(2) & 0.068(2) \\
\hline
\end{tabular}
\end{center}
\caption*{\it Values of the $f^N_{T_{q,G}}$ for the proton and neutron obtained using our estimates of $\Sigma_{\pi N}$
and $\sigma_s$ as described in the text, assuming $z = 1.49$. The values for the heavy quarks are
obtained from those for light quarks and gluons via an ${\cal O}(\alpha_s^3)$ calculation in perturbative QCD.
}
\end{table}

We have also considered the impact of recent calculations of the heavy-quark scalar density matrix elements
$\langle N | {\bar c} c, {\bar b} b, {\bar t} t | N \rangle$. The spread in lattice and phenomenological estimates of
$\sigma_c = m_c \langle N | {\bar c} c | N \rangle$ is quite large, and potentially a large source of uncertainty in
$\sigma_{SI}$. That said, the possible range of $\sigma_c$ includes the value found to ${\cal O}(\alpha_s^3)$
in QCD perturbation theory. The values of the $f^N_{T_{c,b,t}}$ that we find using the ${\cal O}(\alpha_s^3)$ perturbative
calculations of $\sigma_c, \sigma_b$ and $\sigma_t$ are shown in the last three columns of Table~\ref{tab:fTvalues}.
Using the full six-quark expression for $\sigma^p_{SI}$, we find an enhancement of the cross 
section compared to the leading-order three-flavour approximation that is about 10\% for
the CMSSM fixed-point benchmark point that we have studied. We note, however, that the uncertainties in the leading-order three-quark
approximation and the ${\cal O}(\alpha_s^3)$ six-flavour calculation overlap: $\sigma^p_{SI} = (1.25 \pm 0.13) \times 10^{-9}$~pb 
(three quarks)vs $\sigma^p_{SI} = (1.38 \pm 0.17) \times 10^{-9}$~pb (six quarks). As already mentioned,
we consider the latter, using the estimates (\ref{95percent}, \ref{95percentss}, \ref{sigmac}) and (\ref{btnumbers}),
to be the best approximation to the spin-independent WIMP scattering cross section currently available. As also mentioned
above, we consider the spin-dependent WIMP scattering cross section to be relatively well understood.

For the future, we look forward to further refinements of calculations of $\Sigma_{\pi N}$ and $\sigma_s$ using first-principles
lattice techniques as well as phenomenological inputs, recalling that these are the dominant sources of uncertainty in
$\sigma_{SI}$, if one accepts the perturbative calculation of $\sigma_c$. We also look forward to more accurate lattice
calculations of $\sigma_c$, so as to check the accuracy of this perturbative calculation. Lattice calculations have
made great progress over the past decade, but improvement is still desirable.

\section*{Acknowledgements}

We would like to thank M. Hoferichter, F. Kahlhoefer, U. Mei{\ss}ner and M. Voloshin for useful discussions.
The work of JE was supported partly by the United Kingdom STFC Grant ST/P000258/1 and
partly by the Estonian Research Council via a Mobilitas Pluss grant.
The work of N.N. was supported by the Grant-in-Aid for Scientific Research
(No.17K14270). The work of K.A.O. was supported in part by
DOE grant DE-SC0011842 at the University of Minnesota.

\end{document}